\documentclass[twocolumn,aps,showpacs,prb,tightenlines,amsmath,amssymb]{revtex4}
\usepackage{graphicx}
\usepackage{amssymb}
\usepackage{slashed}
\usepackage{dcolumn}
\usepackage{amsmath}
\usepackage{bm}
\usepackage{colordvi}
\usepackage{mathrsfs}
\makeatletter

\newcommand{\Rmnum}[1]{\expandafter\@slowromancap\romannumeral #1@}
\makeatother

\begin{document}

\title{Diamagnetic property and optical absorption in conventional superconductors with magnetic impurities}

\author{F. Yang}
\email{yfgq@mail.ustc.edu.cn.}

\affiliation{Hefei National Research Center for Physical Sciences at the Microscale, Department of Physics, and CAS Key Laboratory of Strongly-Coupled
Quantum Matter Physics, University of Science and Technology of China, Hefei,
Anhui, 230026, China}

\author{M. W. Wu}
\email{mwwu@ustc.edu.cn.}

\affiliation{Hefei National Research Center for Physical Sciences at the Microscale, Department of Physics, and CAS Key Laboratory of Strongly-Coupled
Quantum Matter Physics, University of Science and Technology of China, Hefei,
Anhui, 230026, China}

\date{\today}

\begin{abstract}

By solving the renormalization of the $s$-$d$ interaction from magnetic impurities embeded in conventional superconductors at low concentration, we derive the macroscopic superconducting phase fluctuation and electromagnetic properties within the path-integral approach. It is found that there exist two superconducting phase modes, both exhibiting similar behaviors of the Nambu-Goldstone mode. The existence of two phase modes suggests that in addition to the conventional free Cooper pairs as in the BCS case, there emerges a small part of the localized Cooper pairs around magnetic impurities due to the quantum correlation by the $s$-$d$ interaction, acting as Josephson islands.
The emerging impurity Shiba bands inside the superconducting gap then correspond to the excitations of the ground state of the localized Cooper pairs, associated with the breaking of these Cooper pairs. In the diamagnetic response, the state of the free Cooper pairs gives rise to the conventional real contribution in the generated supercurrent, whereas the one of the localized Cooper pairs results in an imaginary contribution, leading to the superconducting Friedel oscillation, i.e., oscillation in the decay of the vector potential in the Meissner effect. As for the optical absorption of a conventional superconductor lying in the anomalous-skin-effect region, it is found that besides the conventional interband transition of Bogoliubov quasiparticles as revealed by Mattis-Bardeen theory, there also exist the interband transition between the impurity Shiba bands as well as all interband transitions between Bogoliubov quasiparticle and impurity Shiba bands. These transitions exhibit clear and separate resonance characters, providing a feasible scheme for the experimental detection.

\end{abstract}


\maketitle 

\section{Introduction}

In the past few decades, the in-gap excitations in superconducting systems have attracted much attention as they share robust gap protection from superconductors and exhibit a long-range phase coherence, allowing for the desired manipulation in potential application. Various proposals have therefore been put forward in the literature, such as the vortex bound state\cite{VB1,VB2} and Yu-Shiba-Rusinov state (YSR)\cite{Yu,Shiba,Ru} in conventional BCS superconductors, Andreev bound state confined in normal region of short Josephson junction\cite{AB1,AB2,AB3,AB4,AB5,AB6} as well as Majorana bound state localized at the boundaries of topological superconductors\cite{MB1,MB2,MB3,MB4,MB5,MB6,MB7}. Among them, the YSR state, which was first analytically revealed by Yu\cite{Yu}, and later by Shiba\cite{Shiba} as well as Rusinov\cite{Ru} in 1960s considering a classical local spin in a conventional BCS superconductor, has recently attracted the renewed and growing interest. This type of state is characterized as a pair of in-gap bound states that appear around single magnetic impurity embedded in an $s$-wave superconductor, with the particle-hole-symmetric excitation energies $\pm\eta\Delta_0$ associated with the local Cooper-pair breaking by magnetism. Here, $0<\eta<1$ and $\Delta_0$ denotes the superconducting gap. Induced by local magnetism, the YSR state as the in-gap excitation is expected to enable the investigation in the Andreev tunneling process\cite{SA} as well as the study of the magnetic phenomena in superconductors\cite{SM1,SM2,SM3,SM4} with high energy resolution. As for the case at finite impurity concentration, Shiba predicted a pair of impurity bands inside the superconducting gap formed by hybridization of the YSR states from individual magnetic impurities\cite{Shiba}, via numerically calculating the self-consistent self-energy within the random phase approximation. Nowadays, thanks to the advanced fabrication technique that can tailor and control down to each individual atom, the band formed by hybridization of the YSR states is predicted to achieve the topological superconductivity\cite{SAP1,SAP2,SAP3,SAP4,SAP5,SAP6,SAP7,SAP8,SAP9} as potential platform for quantum computational architectures.

Inspired by the renewed attention, a great deal of experimental efforts have been devoted to the search for the existence of the YSR state. The scanning tunneling microscopy and spectroscopy (STM/STS) techniques have been widely applied in the literature to identify a pair of the in-gap resonance peaks that are symmetrically located around zero-bias. Such observation was first reported on the surface of superconducting Nb sample with Mn and Gd adatoms\cite{STM0}, and have now been observed in a variety of systems, ranging from different magnetic adatoms\cite{STMA1,STMA2,STMA3,STMA4,STMA5,STMA6,STMA7,STMA8}, magnetic molecules\cite{SM2,STMM1,STMM2,STMM3,STMM4,STMM5}, magnetic nanostructures (such as ferromagnetic nanowires\cite{SAP4} or artificial atomic chains\cite{SAP6}) and magnetic islands\cite{STMI1,STMI2} embedded in conventional superconductors, over magnetic molecular junctions with proximity-induced superconductivity\cite{STMP}, to iron-based unconventional superconductors\cite{STMHT1,STMHT2,STMHT3}. In comparison with the tremendous experimental progress for the YSR state around single magnetic impurity, the scheme to detect the impurity Shiba band at finite impurity concentration is still absent in the literature, since such detection requires the macroscopic measurements concerning the non-equilibrium properties, which are beyond the STM/STS technique. 

For elucidating the physics of superconductivity and exploring the novel properties, the electromagnetic responses have played a significant role in the past. On one hand, the diamagnetic effect caused by induced supercurrent in magnetic response, referred to as Meissner effect\cite{Meissner,London}, is known as one of the fundamental phenomena of superconductors. On the other hand, the optical spectroscopy in superconductors has been proved as powerful tool to access the properties of superconductivity. Particularly, for superconductors lying in the anomalous-skin-effect region with smaller skin depth compared with the mean free path\cite{NSL0,GIKE3}, by measuring the optical conductivity $\sigma_s(\Omega)=\sigma_{1s}(\Omega)+i\sigma_{2s}(\Omega)$ in linear optical response, fitting $1/\Omega$-like divergent behavior in the imaginary part $\sigma_{2s}(\Omega)$ at low-frequency regime gives rise to the density of the superfluid\cite{Infrad2,Infrad3,MS1,MS3,THZ3,THZ4,NSB2}. The real part $\sigma_{1s}(\Omega)$ (i.e., optical absorption) around terahertz-frequency regime at $T=0~$K is attributed to the interband transition of Bogoliubov quasiparticles, according to the Mattis-Bardeen theory\cite{MB,MBo}. Thus, in $s$-wave superconductors at $T=0~$K, $\sigma_{1s}(\Omega)$ vanishes at $\Omega<2\Delta_0$ but becomes finite for $\Omega$ above $2\Delta_0$\cite{MB}, providing a clear feature to measure the value and symmetry of the superconducting gap\cite{L1,L2,L3,L4,L5,L6,L7,L8,NSL0}. Consequently, as the YSR state is assocciated with the local Cooper-pair breaking, finit-concentration magnetic impurities embededin conventional superconductors is expect to influence the superfluid density as well as optical absorption. By studying this influence, one can therefore reveal the feasible scheme to detect the impurity Shiba bands, and gain a deeper understanding of the competition/coexistence of the superconductivity and local magnetism, which has been one of the focus and intriguing topics in the field. 

In this work, in conventional superconductors with magnetic impurities at low concentration, by analytically solving the renormalization of the $s$-$d$ interaction, we derive the superconducting phase fluctuation and electromagnetic properties within the path-integral approach. Specifically, to elucidate the macroscopic physical picture of the ground state behind the emerging impurity Shiba bands, we calculate the superconducting phase fluctuation within the path-integral approach.  It is found that there exist two superconducting phase modes, and both become inactive after the coupling with the long-range Coulomb interaction which causes the original gapless spectra lifted up to the high-energy frequency as a consequence of Anderson-Higgs mechanism\cite{AHM}, similar to the conventional Nambu-Goldstone mode\cite{Gm1,Gm2,gi0,AK,Ba0,Am0}.
As the existence of the collective phase mode is a direct consequence of the formation of the superconducting state due to the spontaneous breaking of the continuous $U(1)$ symmetry\cite{gi0,Gm1,Gm2,AK,Ba0,Am0,GIKE2}, two phase modes  in superconductors with magnetic impurities suggests that there exist two types of states of the Cooper pairs, forming the ground state through the direct product:  a small part of the Cooper pairs become localized around individual magnetic impurities due to the quantum correlation by the $s$-$d$ interaction, acting
as Josephson islands, similar to the
case of the granular superconductors\cite{GS1,GS2}; the remaining part of the Cooper pairs is still conventional free one as in the BCS case.  The Bogoliubov quasiparticle continuum and emerging impurity Shiba bands then correspond to the excitations of the ground states of the free and localized Cooper pairs, associated with the corresponding pair breaking, respectively.

\begin{figure}[htb]
  {\includegraphics[width=8.2cm]{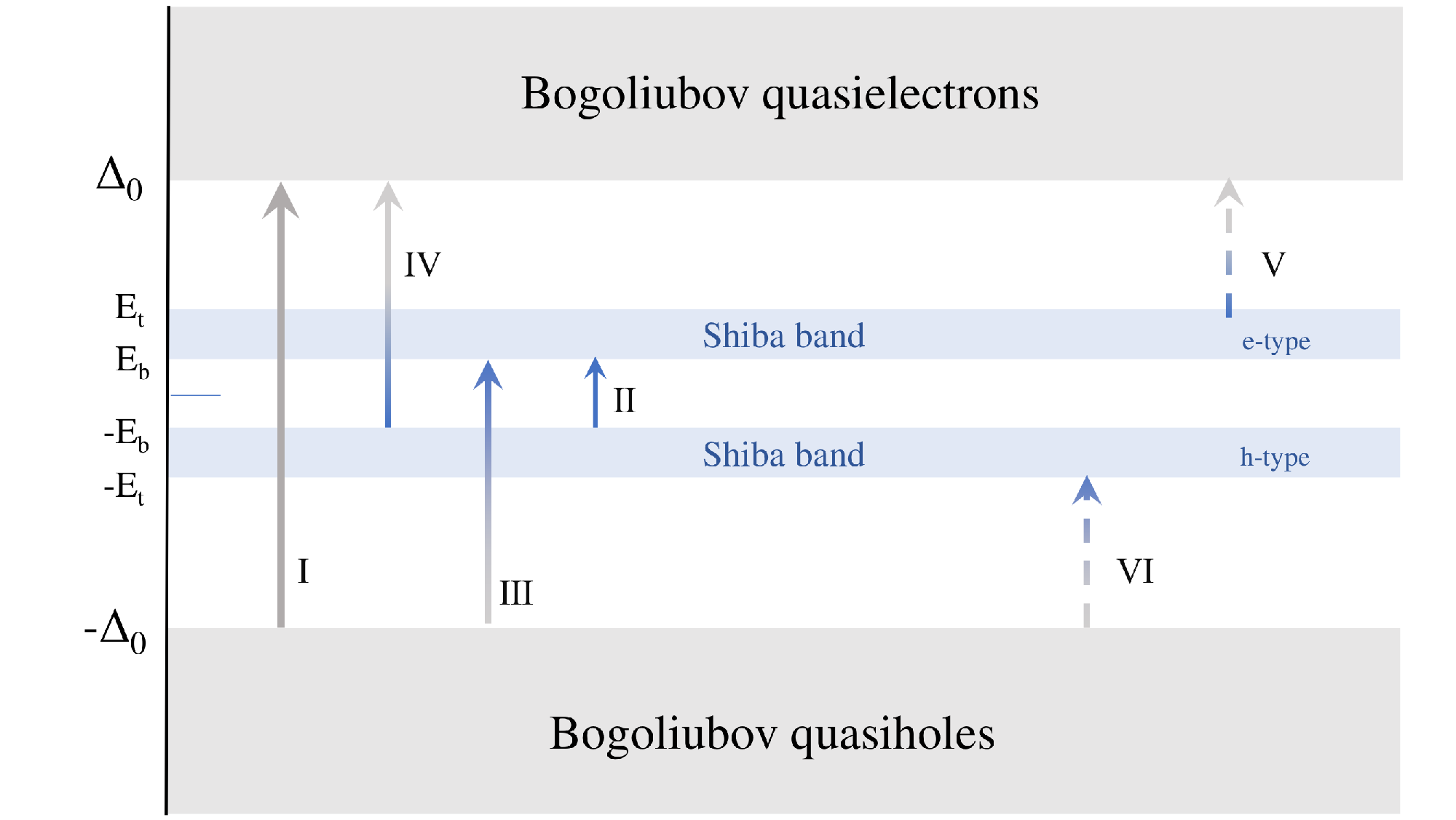}}
  \caption{Interband transitions in conventional superconductors with magnetic impurities. In the figure, the blue and gray shadow regions denote the impurity Shiba bands and Bogoliubov quasiparticle continuum, respectively; $E_t$ and $E_b$ stand for the top and bottom edges of the electron-type (e-type) impurity Shiba band, respectively; the dashed arrow denotes the interband transition from electron type to electron type or from hole type to hole type that occurs at non-zero temperature; the solid arrow represents the interband transition from hole type to electron type that can occur at zero temperature. }    
\label{figyw1}
\end{figure}

The proposed picture of the ground state with free and localized Cooper pairs can well capture the derived electromagnetic properties in the linear response.  On one hand, in the diamagnetic response, the state of the free Cooper pairs gives rise to the conventional real contribution in the generated supercurrent as in the BCS case, whereas the one of the localized Cooper pairs results in an imaginary contribution, which can be understood by the $\pi/2$-phase difference between wave-vectors of the free and localized Cooper pairs. Consequently, in the diamagnetic response, in contrast to the
exponential decay in the conventional Meissner effect\cite{London,G1}, the imaginary contribution in supercurrent due to the $s$-$d$ interaction from magnetic impurities leads to an oscillation in the decay of the vector potential from the surface to the interior of superconductors, similar to the Friedel oscillation in normal metals\cite{MH} due to the local modulation of the charge density by defect. We therefore refer to this oscillation in superconductors with magnetic impurities as superconducting Friedel oscillation.
On the other hand, it is noted that the impurity Shiba bands and Bogoliubov quasiparticle continuum, as corresponding excitations of the ground states of the localized and
free Cooper pairs, are similar to each other. Because of this similarity, 
in the optical absorption of a conventional $s$-wave superconductor lying in the anomalous-skin-effect region\cite{NSL0,GIKE3}, at zero temperature, besides the conventional interband transition (channel I in Fig.~\ref{figyw1}) of Bogoliubov quasiparticles as revealed by Mattis-Bardeen theory\cite{MB}, there also exist the interband transitions (from hole type to electron type) between the impurity Shiba bands  (channel II in Fig.~\ref{figyw1}) as well as between Bogoliubov quasiparticle and impurity Shiba bands (channels III and IV in Fig.~\ref{figyw1}).  The channel II leads to a resonance peak centered around $\Omega=2\eta\Delta_0$ in optical spectroscopy, whereas channles III and IV cause a crossover at $\Omega=\Delta_0+E_b$, with $E_b$ being the bottom edge of the impurity Shiba band. Interestingly, with increase of temperature from zero, between Bogoliubov quasiparticle and impurity Shiba bands, there gradually emerge the interband transitions from electron (hole) type to electron (hole) type, as shown by channel V (VI) in Fig.~\ref{figyw1}, leading to a crossover at $\Omega=\Delta_0-E_t$, with $E_t$ being the top edge of the impurity Shiba band. Consequently, a feasible scheme for experimental detection of the impurity Shiba band is proposed, by measuring the emerging character in diamagnetic property and/or optical spectroscopy.

\section{Model}

In this section, we first introduce the Hamiltonian of superconductors in the presence of the $s$-$d$ interaction between electrons and magnetic impurities, and show the corresponding renormalized Green function revealed by Shiba at finite impurity concentration within the random phase approximation. In contrast to the numerical formulation by Shiba, we present the analytical solution of the complex renormalization by $s$-$d$ interaction in the Green function at a low concentration of the magnetic impurities. A finite density of states which is centered around $\eta\Delta_0$ with bandwidth proportional to the square root of the impurity concentration emerges inside the superconducting gap, suggesting the emergence of the impurity Shiba band as revealed by numerical calculation from Shiba. Then, using the analytically obtained Green function, we present the diagrammatic formalism within the path-integral approach to investigate the electromagnetic properties of superconductors in the linear regime. 

\subsection{Hamiltonian and renormalized Green function}
\label{secHA}

In conventional $s$-wave superconductors, the total Hamiltonian with the $s$-$d$ interaction between electrons and magnetic impurities in Nambu$\otimes$spin space reads\cite{Shiba}
\begin{eqnarray}
  H&=&\frac{1}{2}\sum_{\bf k}\psi^{\dagger}_{\bf k}(\xi_{\bf k}\tau_3\!-\!\Delta_0\tau_2\sigma_2)\psi_{\bf k}\!-\!\frac{1}{2}J\sum_{\bf kk'}\psi^{\dagger}_{\bf k}\tilde{\bm \sigma}\psi_{\bf k'}\!\cdot\!{\bf S},~~~~\label{Ham}
\end{eqnarray}
where the field operator $\psi_{\bf k}=(\psi_{{\bf k}\uparrow},\psi_{{\bf k}\downarrow},\psi^{\dagger}_{-{\bf k}\uparrow},\psi^{\dagger}_{-{\bf k}\downarrow})^{T}$; $\xi_{{\bf k}}=k^2/(2m)-\mu$ with $m$ denoting the effective mass and $\mu$ being the chemical potential; $\sigma_i$ and $\tau_i$ are the Pauli matrices in spin and Nambu particle-hole space, respectively; ${\tilde{\bm \sigma}}={\bm \sigma}{(1+\tau_3)/}{2}+\sigma_2{\bm \sigma}\sigma_2(1-\tau_3)/{2}$; 
${\bf S}$ and $J$ denote the local spin and exchange interaction in the $s$-$d$ interaction, respectively. Particularly, in consideration of a classical spin, one has $S_x^2=S_y^2=S_z^2=S^2/3$ and $S_xS_y=S_xS_z=S_yS_z=0$.

It is established that the Green-function formalism provides an efficient approach to elucidate the single-particle excitation spectrum. In general, the Green function is defined as $G_{\bf k}(\omega)=-i\langle\psi_{\bf k}(\omega)\psi_{\bf k}^{\dagger}(\omega)\rangle$, which can be solved through the Dyson equation\cite{G1,MH}:
\begin{equation}\label{Greenfunction}
G_{\bf k}(\omega)=G_{0{\bf k}}(\omega)+G_{0{\bf k}}(\omega)\Sigma(\omega)G_{\bf k}(\omega).  
\end{equation}
Here, $G_{\bf k}^{(0)}(\omega)$ represents the bare Green function and $\Sigma(\omega)$ denotes the self-energy due to the external interaction. The bare Green function of the conventional BCS superconductors is established as\cite{G1} 
\begin{equation}\label{G0}
G_{0{\bf k}}(\omega)=\frac{\omega\!+\!\xi_{\bf k}\tau_3\!-\!\Delta_0\tau_2\sigma_2}{\omega^2-\xi_{\bf k}^2-\Delta_0^2},  
\end{equation}
and within the random phase approximation to take random spatial distribution and random orientation of individual local spins, the self-energy due to the $s$-$d$ interaction between electrons and magnetic impurities is given by\cite{Shiba}
\begin{equation}\label{SE}
\Sigma(\omega)=n_i({\bf S}\cdot{\tilde{\bm \sigma}})Z(\omega)({\bf S}\cdot{\tilde{\bm \sigma}})+({\bf S}\cdot{\tilde{\bm \sigma}})Z(\omega)\Sigma(\omega)Z(\omega)({\bf S}\cdot{\tilde{\bm \sigma}}),  
\end{equation}  
where $Z(\omega)=\sum_{\bf k}G_{\bf k}(\omega)$ and $n_i$ represents the impurity concentration. 

To self-consistently calculate the Green function from Eqs.~(\ref{Greenfunction}) and (\ref{SE}), based on the bare one in Eq.~(\ref{G0}), one can consider a renormalized Green function as\cite{G1,Shiba} 
\begin{equation}\label{RG}
G_{\bf k}(\omega)=\frac{\tilde\omega\!+\!\xi_{\bf k}\tau_3\!-\!\tilde\Delta_0\tau_2\sigma_2}{\tilde\omega^2-\xi_{\bf k}^2-\tilde\Delta_0^2},  
\end{equation}
and the self-energy in Eq.~(\ref{SE}) becomes
\begin{equation}\label{FSE}
\Sigma(\omega)=\frac{n_i(JS/2)^2Z(\omega)}{1-[JSZ(\omega)/2]^2}.  
\end{equation}
Then, substituting Eqs.~(\ref{RG})-(\ref{FSE}) into Eq.~(\ref{Greenfunction}), one arrives at the renormalization equations revealed by Shiba\cite{Shiba}:
\begin{equation}\label{RE}
\frac{\omega}{\Delta_0}=\frac{\tilde\omega}{\tilde\Delta_0}\bigg[1\!-\!\frac{\gamma_s}{\Delta_0}\frac{\sqrt{1\!-\!(\frac{\tilde\omega}{\tilde\Delta_0})^2}}{\eta^2\!-\!(\frac{\tilde\omega}{\tilde\Delta_0})^2}\bigg],  
\end{equation}
and
\begin{equation}\label{RD}
{\tilde\Delta_0}=\Big[1-\frac{1-(JSD{\pi}/2)^2}{2}\Big(1-\frac{{\omega}/{\Delta_0}}{{\tilde\omega}/{\tilde\Delta_0}}\Big)\Big]\Delta_0.
\end{equation}
Here, $\gamma_s=2n_iD\pi(JS/2)^2/[1+(JSD{\pi}/2)^2]^2$ denotes the relaxation rate due to the $s$-$d$ interaction; $D$ denotes the density of states at the Fermi level in the normal state; the coefficient $\eta$ is written as 
\begin{equation}
\eta=\frac{1-(JSD{\pi}/2)^2}{1+(JSD{\pi}/2)^2},
\end{equation}
which is related to the energies $\pm\eta\Delta_0$ of the pair of the YSR state around single magnetic impurity in a conventional $s$-wave superconductor\cite{Yu,Shiba,Ru}.

As the imaginary part of the $\sigma_0\tau_0$ component of the retarded Green function corresponds to the spectra function, one can calculate the density of states as
\begin{equation}
\rho(\omega)={\rm Im}{\rm Tr}[Z(\omega+i0^+)/4]=-{\rm Im}\Big[\frac{\pi{D}\tilde\omega}{\sqrt{\tilde\Delta_0^2-\tilde\omega^2}}\Big].\label{DOE}  
\end{equation}
Without magnetic impurities ($\gamma_s=0$) and hence the renormalization according to Eqs.~(\ref{RE}) and (\ref{RD}), the density of states $\rho(\omega)$ from Eq.~(\ref{DOE}) becomes finite at $\omega\ge\Delta_0$ but vanishes for $0<\omega<\Delta_0$ as it should be, since the continuum of the Bogoliubov quasiparticle lies above the superconducting gap.

In the presence of the magnetic impurities, as seen from Eq.~(\ref{RE}), there exist the complex solutions of the renormalization ${\tilde\omega}/{\tilde\Delta_0}$ when $\omega>\Delta_0$, suggesting the existence of the interaction between Bogoliubov quasiparticles and magnetic impurities. Particularly, a further numerical calculation of Eq.~(\ref{RE}) by Shiba\cite{Shiba} revealed that there exists additional complex solutions of ${\tilde\omega}/{\tilde\Delta_0}$ when $0<\omega<\Delta_0$, and this complex renormalization leads to a finite density of states $\rho(\omega)$ [Eq.~(\ref{DOE})] inside the superconducting gap, suggesting the emergence of the impurity Shiba band.  It is also revealed\cite{Shiba} that the density of states of the emerging electron-type impurity Shiba band is centered around $\eta\Delta_0$ with bandwidth proportional to the square root of the impurity concentration, whereas in consideration of the particle-hole symmetry, a corresponding hole-type impurity Shiba band emerges symmetrically at $\omega<0$.

\subsection{Solution of renormalization at low concentration}
\label{sec-as}

For the formulation of the electromagnetic properties within the Green-function formalism, the numerical results of the impurity Shiba bands are hard to handle for the practical calculation. In this part, we present the analytical solution of Eq.~(\ref{RE}) at low impurity concentration with small dimensionless ratio $r=\gamma_s/\Delta_0$.

By defining $x={\omega}/{\Delta_0}$, we consider a complex solution of the renormalization:
\begin{equation}\label{as}
{\tilde\omega}/{\tilde\Delta_0}=x+\delta{x}+im,  
\end{equation}
in which the real parameters $\delta{x}$ and $m$ are small quantities for weak renormalization at low impurity concentration.

For the branch of the solutions of the impurity Shiba bands at $\omega>0$, considering the fact that the narrow impurity Shiba band is away from the edge of the continuum of Bogoliubov quasiparticle, Eq.~(\ref{RE}) can be written as
\begin{equation}\label{ISIS1}
\delta{x}+im~{\approx}~r\frac{(x+\delta{x}+im)(\sqrt{1-x^2}-\frac{imx}{\sqrt{1-x^2}})}{\eta^2-(x+\delta{x}+im)^2}.  
\end{equation}
From above equation, one can analytically derive the solutions of the renormalization (refer to Appendix~\ref{sec-a1}):
\begin{eqnarray}
  m^2=-B(x)+\sqrt{[B(x)]^2+rW(x)-(\eta^2-x^2)^2},  \label{mS}
\end{eqnarray}
and
\begin{equation}\label{R1}
  \delta{x}=\frac{rx\sqrt{1-x^2}(\eta^2-x^2+m^2)}{(\eta^2-x^2+m^2)^2+4m^2x^2}.
\end{equation}  
Here, $B(x)=\eta^2+x^2-\frac{r/2}{\sqrt{1-x^2}}$ and $W(x)=\sqrt{1-x^2}(\eta^2+x^2)-\frac{x^2(\eta^2-x^2)}{\sqrt{1-x^2}}$. It is noted from Eq.~(\ref{mS}) that the imaginary part of the renormalization has defined solutions only in the regime with $m^2\ge0$, which limits the energy regime $[E_b,E_t]$ of the emerging density of states inside the superconducting gap and hence the impurity Shiba band. Mathematically, since the condition of $m^2\ge0$ prefers a low factor $(\eta^2-x^2)^2$ in Eq.~(\ref{mS}), the solution of $m$ is centered around $\eta\Delta_0$, whereas the factor $rW(x)$ determines the bandwidth $\Delta{E}=E_t-E_b$ of the solution, as one requires $rW(x)\ge(\eta^2-x^2)^2$ for condition of $m^2\ge0$ in Eq.~(\ref{mS}). Particularly, from Eq.~(\ref{mS}), by solving $m^2=0$ and hence $rW(x)=(\eta^2-x^2)^2$ for $x>0$, at low concentration of the magnetic impurities, one has the solutions:
\begin{eqnarray}
  x_{m=0}&=&[\eta^2\pm\sqrt{rW(x)}]^{1/2}\approx[\eta^2\pm\sqrt{rW(\eta)}]^{1/2}\nonumber\\
  &=&\eta\pm{\sqrt{2r\sqrt{1-\eta^2}}}/{2},
\end{eqnarray}
which correspond to the top and bottom edges of the energy spectrum of the impurity Shiba band (i.e., $E_t/\Delta_0$ and $E_b/\Delta_0$). Therefore, with $(E_t+E_b)/2=\eta\Delta_0$ and {\small $\Delta{E}=\Delta_0\sqrt{2r\sqrt{1-\eta^2}}$}, the density of states of the impurity Shiba band is centered around $\eta\Delta_0$ with bandwidth proportional to the square root of the impurity concentration as well as to the factor $(1-\eta^2)^{1/4}$.  Particularly, with $n_i\rightarrow0^+$ and hence the vanishing hybridization of the YSR state, for $m^2\ge0$, one finds the sole solution of $x=\eta$ in Eq.~(\ref{mS}), which corresponds to the YSR state around single magnetic impurity. All these characters from our analytical derivation agree well with the ones from the calculation by Shiba\cite{Shiba}. 

As for the branch of the solutions of the continuum of the Bogoliubov quasiparticle, which is away from the narrow impurity Shiba bands at low impurity concentration, 
Eq.~(\ref{RE}) can be written as
\begin{equation}\label{IQIQ1}
\delta{x}+im~{\approx}~r\frac{(x+\delta{x}+im)\sqrt{1-(x+\delta{x}+im)^2}}{\eta^2-x^2},  
\end{equation}
and then, considering the weak renormalization, the 
solutions of the renormalization read (refer to Appendix~\ref{sec-a1})
\begin{equation}\label{mSQ}
m^2=\frac{{r^2x^2(x^2-1)}/{[(\eta^2-x^2)^2]}}{1+r^2x^2/(\eta^2-x^2)^2},  
\end{equation}
and
\begin{equation}\label{R2}
  \delta{x}=\frac{rx\sqrt{{|1-x^2+m^2-2imx|+1-x^2+m^2}}}{\sqrt{2}(\eta^2-x^2)}.
\end{equation}  
In this situation, the imaginary part has defined solutions only in the regime with $x\ge1$ and hence $m^2\ge0$ as it should be, since the continuum of the Bogoliubov quasiparticle lies above the superconducting gap. 

In other regimes ($x<E_b/\Delta_0$ and $E_t/\Delta_0<x<1$), one has the vanishing imaginary part (i.e., $m=0$), and in this situation, at low impurity concentration with small $r$, the solution of the renormalization reads
\begin{equation}\label{R3}
\delta{x}=\frac{rx\sqrt{1-x^2}}{(\eta^2-x^2)}.  
\end{equation}
It is noted that the solution of the real part $\delta{x}$ of the renormalization in Eqs.~(\ref{R1}), (\ref{R2}) and (\ref{R3}) is analytically continuous at the boundaries with $m=0$, guaranteeing the analytic continuity of the derived solution in the entire energy regime for the practical calculation.   

Consequently, in contrast to the numerical formulation by Shiba, we obtain the analytical solutions of the complex renormalization ${\tilde\omega}/{\tilde\Delta_0}$ [Eq.~(\ref{as})], and then, the renormalized gap $\tilde\Delta_0$ [Eq.~(\ref{RD})] as well as the renormalized Green function in Eq.~(\ref{RG}) and density of states in Eq.~(\ref{DOE}) can be obtained for the practical formulation.

\subsection{Diagrammatic formalism for electromagnetic properties in linear regime}

In this part, in the presence of the $s$-$d$ interaction from magnetic impurities at finite concentration, we present the diagrammatic formalism for the electromagnetic properties of superconductors in the linear response. Specifically, considering the presence of the vector potential ${\bf A}$ and long-range Coulomb interaction for the formulation of the electromagnetic properties, with the $s$-$d$ interaction between electrons and magnetic impurities, the action of an $s$-wave superconductor after the Hubbard-Stratonovich transformation is written as\cite{Ba0,G1}
\begin{eqnarray}
  \!\!\!S&=&\!\!\!\int{dx}\bigg\{\sum_{s=\uparrow,\downarrow}\!\!\psi^*_s(x)[i\partial_{t}\!-\!\xi_{\hat {\bf p}-e{\bf A}}-\mu_H(x)]\psi_s(x)\nonumber\\
  &&\mbox{}\!\!\!+\psi^{\dagger}(x)[\Delta(x)\tau_+-\Delta^*(x)\tau_-]i\sigma_2\psi(x)\!-\!\frac{|\Delta(x)|^2}{U}\nonumber\\
  &&\mbox{}\!\!\!+J\!\sum_{ss'}\psi_s^*(x){\bm \sigma}_{ss'}\psi_{s'}(x)\!\cdot\!{\bf S}\bigg\}\!+\!\int\!{dt}d{\bf q}\frac{|\mu_{H}({\bf q})|^2}{2V_{\bf q}}.~~~~ \label{SC}
\end{eqnarray}
Here, the superconducting order parameter reads $\Delta(x)=[\Delta_{0}+\delta|\Delta|(x)]e^{i\delta\theta(x)}$ with $\delta|\Delta|$ and $\delta\theta(x)$ being the amplitude and phase fluctuations, respectively; the momentum operator ${\hat {\bf p}}=-i\hbar{\bm \nabla}$; $\mu_H$ denotes the Hartree field that is related to long-range Coulomb interaction\cite{Ba0}; $U$ represents the BCS pairing potential and $V_{\bf q}=e^2/(q^2\epsilon_0)$ stands for the Fourier component of the Coulomb potential.

In Nambu$\otimes$spin space, using the unitary transformation:
\begin{equation}
  \psi(x){\rightarrow}e^{i\tau_3\delta\theta(x)/2}\psi(x),
\end{equation}
the action becomes 
\begin{eqnarray}
S&=&\frac{1}{2}\int{dx}\bigg\{\psi^{\dagger}(x)\big[G_0^{-1}({\hat {p}})\!-\!{\hat \Gamma}\big]\psi(x)\!-\!\eta_f(\frac{p_s^2}{2m}+\mu_{\rm eff})\nonumber\\
&&\mbox{}-\!\frac{|\Delta(x)|^2}{U}\bigg\}\!+\!\int\!{dt}d{\bf q}\frac{|\mu_{H}({\bf q})|^2}{2V_{\bf q}}, \label{BdGaction}
\end{eqnarray}
with the vertex operator:
\begin{equation}
{\hat \Gamma}=\frac{{\bf p}_s\cdot{\hat{\bf p}}}{m}+\frac{p^2_s}{2m}\tau_3+\delta|\Delta|\sigma_2\tau_2-J{\bf S}\cdot{\tilde{\bm \sigma}}+\mu_{\rm eff}\tau_3.  
\end{equation}
Here, $G_{0{\hat p}}^{-1}=i\partial_{t}-\xi_{\hat {\bf p}}-{\Delta}_0\sigma_2\tau_2$ and $\eta_f=\sum_{\bf k}2$ emerges because of the anti-commutation of Fermi field; the superconducting momentum ${\bf p}_s=\nabla\delta\theta/2-e{\bf A}$ and effective field $\mu_{\rm eff}={\partial_t\delta\theta}/{2}+\mu_{H}$. It is noted that the first term in vertex operator ${\hat \Gamma}$ is related to the current vertex ${\bf {\hat p}}/m$ and drives a current (drive effect) through the conventional current-current correlation\cite{G1,MH,DDT,PT1}, whereas the second term in ${\hat \Gamma}$ is related to the density vertex $\tau_3$ and can directly pump a current $-e^2n{\bf A}/m$ (pump effect), which is conventionally considered as an unphysical non-gauge-invariant current in the literature\cite{G1,MH,PT1}.

Then, through the standard integration over the Fermi field within the path-integral approach, one obtains the effective action: 
\begin{eqnarray}
  {S}&=&\!\!\!\frac{1}{2}\int\!{dx}\Big\{-i{\rm {\bar Tr}}\ln[G_0^{-1}]\!+\!i{\rm{\bar Tr}}\sum_{n=1}^{\infty}\frac{1}{n}[(G_0\Gamma)^n]\!-\!\frac{|\Delta|^2}{U}\nonumber\\
    &&\mbox{}-\eta_f\Big(\mu_{\rm eff}+\frac{p_s^2}{2m}\Big)\Big\}\!+\!\int\!{dt}d{\bf q}\frac{|\mu_{H}({\bf q})|^2}{2V_{\bf q}}.\label{sss}
\end{eqnarray}
In the present work, we focus on the influence of the impurity Shiba band on the linear response of superconductors. In this situation, one only needs to keep the second order of the external field but all orders of the $s$-$d$ interaction, whereas the odd orders of the $s$-$d$ interaction vanishes within the random phase approximation\cite{Shiba}. In addition, it is well established that being charge neutral and spinless, the amplitude fluctuation, known as the Higgs mode of superconductors in the literature\cite{Am0,OD1,OD2,OD3,Am12,Am6,symmetry}, does not responds to the electromagnetic field in the linear regime at long-wavelength limit\cite{GIKE2,GIKE4,PT2,PT3}, i.e., $\delta|\Delta|=0$.

\begin{figure}[htb]
  {\includegraphics[width=8.2cm]{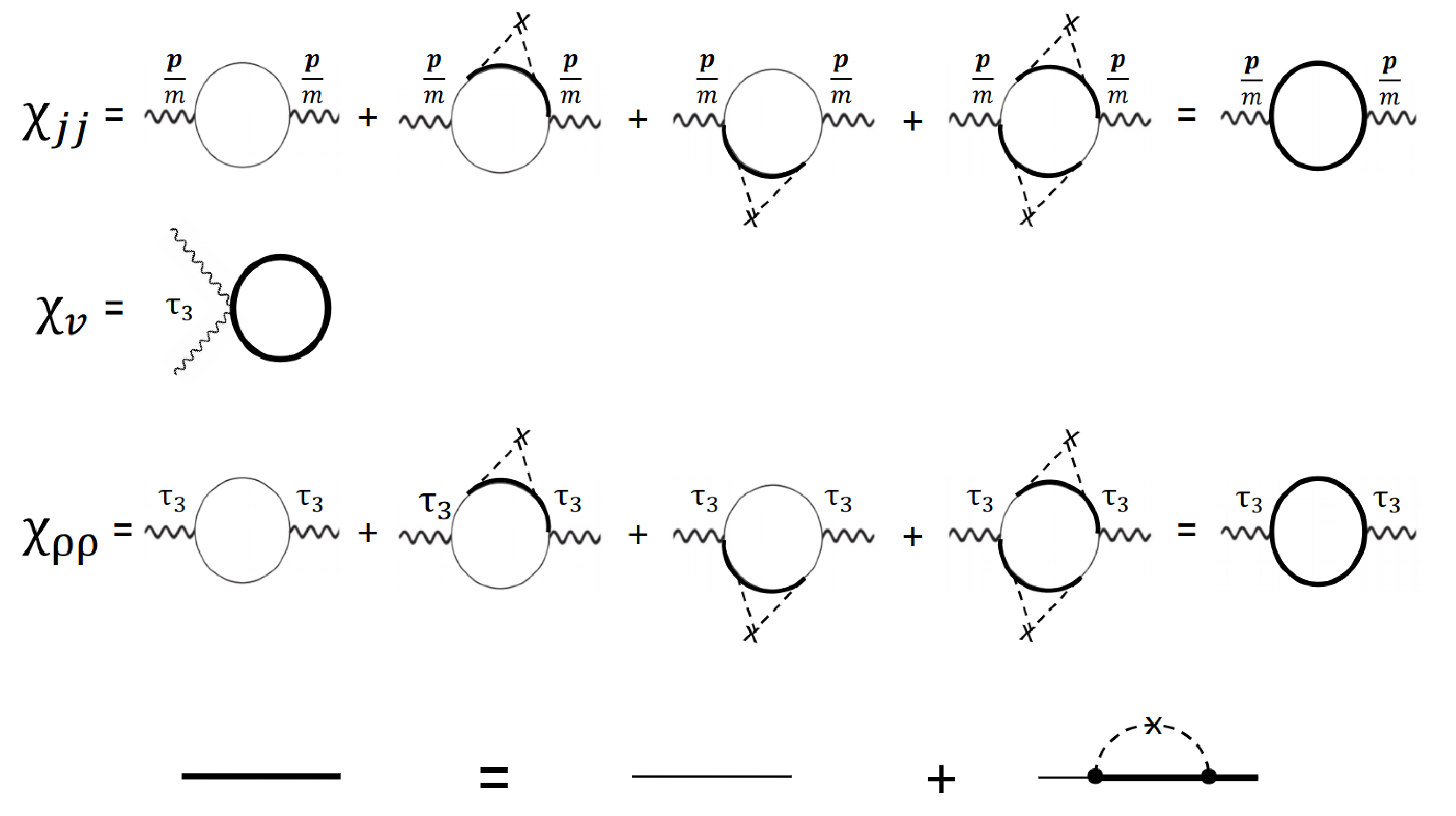}}
  \caption{Diagrammatic formalism of the corresponding correlation coefficients in non-equilibrium action in Eq.~(\ref{naction}). In the figure, the dashed line with cross represents the $s$-$d$ interaction; the wavy line is associated with the external field; the thin and thick solid lines denote the bare and renormalized Green functions, respectively. }    
\label{figyw2}
\end{figure}

Therefore, the non-equilibrium effective action reads 
\begin{equation}
\delta{S}\!=\!\!\int{dt}d{\bf q}\frac{|\mu_{H}|^2}{2V_{\bf q}}-\!\int\!{dx}\Big[\frac{\chi_{v}+{\chi}_{jj}}{2}\frac{p_s^2}{2m}-\mu_{\rm eff}^2{\chi_{\rm \rho\rho}}\Big],\label{naction} 
\end{equation}
where the first-order correlation due to the direct density-vertex contribution by pump effect:
\begin{equation}
  \chi_{v}=\eta_f-\sum_{n=0}^{\infty}i{\rm {\bar Tr}}[G_0\tau_3(G_0J{\bf S}\cdot{\tilde{\bm \sigma}})^{2n}]=\eta_f-i{\rm {\bar Tr}}(G\tau_3),  \label{VT}
\end{equation}
and the second-order current-current correlation because of the drive effect:
\begin{eqnarray}
  \chi_{jj}&=&-i{\rm {\bar Tr}}\Big\{\frac{k^2}{3m}(G_0\tau_0)^2[1\!+\!2(G_0J{\bf S}\!\cdot\!{\tilde{\bm \sigma}})^2\!+\!O(J^{n>2})]\Big\}\nonumber\\
  &\approx&-i{\rm {\bar Tr}}\big[\frac{k^2}{3m}(G\tau_0)^2\big],~ \label{jj}
\end{eqnarray}
as well as the second-order density-density correlation due to the effective field $\mu_{\rm eff}$:
\begin{eqnarray}
\chi_{\rm \rho\rho}&=&\frac{i{\rm {\bar Tr}}}{4}\Big\{(G_0\tau_3)^2[1\!+\!2(G_0J{\bf S}\!\cdot\!{\tilde{\bm \sigma}})^2+O(J^{n>2})]\Big\}\nonumber\\
  &\approx&\frac{i{\rm {\bar Tr}}}{4}\big[(G\tau_3)^2\big].~\label{rhorho}
\end{eqnarray}
Here, to consider the influence of the impurity Shiba bands, we have neglected the non-equilibrium vertex and cross-diagram corrections\cite{G1,MH,RS} by $s$-$d$ interaction in both current-current and density-density correlations, and only kept the self-consistent Born corrections (i.e., self-consistent complex renormalization by $s$-$d$ interaction) as the renormalized equilibrium Green function in Dyson equation [Eq.~(\ref{Greenfunction})]\cite{G1,MH}. As for the direct density-vertex contribution in Eq.~(\ref{VT}), the $s$-$d$ interaction provides the renormalization to the fermion bubble (i.e., bare Green function), exactly same as the one by the equilibrium self-energy $\Sigma(\omega)$ in Eq.~(\ref{Greenfunction}). Moreover, in the second order of the external field, the coupling term between current-vertex related ${\bf p}_s\cdot{\hat {\bf p}}/m$ and density-vertex-related $\mu_{\rm eff}\tau_3$ vanishes as a consequence of the particle-hole symmetry. 

Then, with the solved renormalized Green function in Sec.~\ref{sec-as}, by calculating the non-equilibrium action $\delta{S}$ in Eq.~(\ref{naction}), one can derive the electromagnetic properties of superconductor in the linear response to study the influence of the $s$-$d$ interaction from magnetic impurities.

\section{Results}

In this section, to elucidate the macroscopic physical picture behind the emerging impurity Shiba bands, we first derive the equation of motion of the superconducting phase fluctuation in the presence of the magnetic impurities. Then, by formulating the non-equilibrium action in consideration of the stationary magnetic and optical responses, we calculate the diamagnetic property and optical absorption, respectively, and study the influence of the impurity Shiba bands on these properties.

\subsection{Superconducting phase fluctuation and physical picture}
\label{pp}

Using the analytically obtained renormalized Green function in Sec.~\ref{sec-as}, we derive the superconducting phase fluctuation in the presence of the magnetic impurities. Specifically, by taking the superconducting momentum ${\bf p}_s=\nabla\delta\theta/2$,
in the center-of-mass momentum space, after the integration over the Hartree field, the non-equilibrium action in Eq.~(\ref{naction}) directly becomes the effective one of the phase fluctuation: 
\begin{equation}
  \delta{S}_{\delta\theta}\!=\!\int\!\frac{dt{d{\bf q}}}{{\varepsilon}_{\bf q}}\Big[{\chi_{\rho\rho}}\Big(\frac{\partial_t\delta\theta}{2}\Big)^2\!-\!\frac{({1\!+\!2V_q\chi_{\rho\rho}}){\eta}_sn}{m}\Big(\frac{i{\bf q}\delta\theta}{2}\Big)^2\Big],  \label{acp}
\end{equation}
where $\varepsilon_{\bf q}=1+2V_{\bf q}\chi_{\rho\rho}$ denotes the dielectric function; $\eta_s=(\chi_v+\chi_{jj})/(2n)$ stands for the ratio of the superfluid density $n_s$ to the electron density $n$.

To consider the case at nonzero temperature, we preform the formulation within the Matsubara representation [$\omega\rightarrow{i\omega_l}=(2l+1)\pi{T}$, $-i{\rm {\bar Tr}}\rightarrow{\rm {\bar Tr}}$], and then, the related correlation coefficients are given by 
  \begin{eqnarray}
    \chi_{\rho\rho}&=&-T\sum_{{\bf k}l}\frac{\rm Tr}{4}[G_{\bf k}(i\omega_l)\tau_3G_{\bf k}(i\omega_l)\tau_3]=-T\sum_{{\bf k}l}\frac{{\rm Tr}}{4}\nonumber\\
    &&\mbox{}\times{[\tau_3\partial_{\xi_{\bf k}}G_{\bf k}(i\omega_l)]}=\sum_{l}\frac{2DT\omega_D}{\tilde\omega_l^2\!+\!\omega_D^2\!+\!\tilde\Delta_0^2},
  \end{eqnarray}
  and
  \begin{eqnarray}
    \eta_sn&=&\!\!\sum_{\bf k}\Big\{2+T\sum_{l}{\rm Tr}\Big[\tau_3G_{\bf k}(i\omega_l)+\frac{k^2}{6m}G^2_{\bf k}(i\omega_l)\Big]\Big\}\nonumber\\
    &=&\!\!T\sum_{l}\!\!\int\!\frac{4{\pi}dk}{(2\pi)^3}{\rm Tr}\Big\{\frac{k^2G^2_{\bf k}}{6m}(i\omega_l)\!-\!\frac{\tau_3k^3\partial_{k}G_{\bf k}(i\omega_l)}{6}\Big\}\nonumber\\
  &=&\!\!\sum_{{\bf k}l}\frac{4k_F^2T\tilde\Delta_0^2/(3m)}{[(i\tilde\omega_l)^2\!-\!\xi_{\bf k}^2\!-\!\tilde\Delta_0^2]^2}=\sum_l\frac{n\pi{T}\tilde\Delta_0^2}{(\tilde\Delta_0^2\!+\!\tilde\omega_l^2)^{3/2}}, \label{esn}
\end{eqnarray}
where $\omega_D$ denotes the Debye frequency. 

Due to the complex renormalization by the $s$-$d$ interaction from magnetic impurities (the complex solution of the renormalization within the Matsubara-frequency representation refers to Appendix~\ref{sec-a2}), from the effective action of the phase fluctuation in Eq.~(\ref{acp}), there emerge two separate equations of motion of the phase modes:
\begin{eqnarray}
  \Big[\partial_t^2+\frac{{\rm Re}(\eta_s)ne^2}{\epsilon_0m}+\frac{{\rm Re}(\eta_s)n}{{\rm Re}{(\chi_{\rho\rho})}m}q^2\Big]\frac{\delta\theta}{2}=0,\label{pm1}\\
   \Big[\partial_t^2+\frac{{\rm Im}(\chi_{\rho\rho}\eta_s)ne^2}{{\rm Im}(\chi_{\rho\rho})\epsilon_0m}+\frac{{\rm Im}(\eta_s)n}{{\rm Im}{(\chi_{\rho\rho})}m}q^2\Big]\frac{\delta\theta}{2}=0.\label{pm2}
\end{eqnarray}
It is noted that both phase modes in Eqs.~(\ref{pm1}) and~(\ref{pm2}) exhibit gapless linear energy spectrum at free case (i.e., without long-range Coulomb interaction), and show gapped energy spectrum at long-wavelength limit after the coupling to the long-range Coulomb interaction as a consequence of the Anderson-Higgs mechanism\cite{AHM}. Hence, both become inactive, and the original global and rigid phase coherence for achieving robust superconductivity in conventional superconductors\cite{DDT} remains even in the presence of the magnetic impurities. Clearly, the phase mode in Eq.~(\ref{pm1}) corresponds to the conventional Nambu-Goldstone mode\cite{gi0,AK,Ba0,Am0,GIKE2}, as this equation of motion in the case without magnetic impurities exactly recovers the one\cite{GIKE2,DDT,AK} as in the BCS case, whereas the one in Eq.~(\ref{pm2}) emerges totally due to the complex renormalization by $s$-$d$ interaction from magnetic impurities.

According to the Goldstone theorem\cite{Gm1,Gm2}, as the existence of the collective gapless phase mode is a direct consequence of the formation of the marcoscopic superconducting state due to the spontaneous breaking of the continuous $U(1)$ symmetry\cite{gi0,Gm1,Gm2,AK,Ba0,Am0,GIKE2}, the emerging two phase modes here suggests that there exist two types of states of the Cooper pairs, forming the ground state in superconductors with magnetic impurities through the direct product. Specifically, with magnetic impurities, a small part of the Cooper pairs become localized around individual magnetic impurities due to the quantum correlation by the $s$-$d$ interaction, acting as Josephson islands and hence leading to the phase mode in Eq.~(\ref{pm2}), similar to the case of the granular superconductors\cite{GS1,GS2}. However, within the random phase approximation that takes random spatial distribution\cite{MH,Shiba}, the emerging Josephson islands by localized Cooper pairs does not manifest themselves explicitly. The remaining part of the Cooper pairs is still conventional free type, resulting in the phase mode in Eq.~(\ref{pm1}). The impurity Shiba bands and Bogoliubov quasiparticle continuum then correspond to the excitations of the ground state of the localized and free Cooper pairs, respectively.

Based on this picture with localized and conventional free Cooper pairs, one can understand the properties of the single-particle energy spectra in superconductors with magnetic impurities. On one hand, due to the small proportion of the localized Cooper pair compared with the free ones, the YSR state around single magnetic impurity exhibits a small (i.e., in-gap) excitation energy $\eta\Delta$ associated with the breaking of the localized Cooper pair, whereas the enhancement of the exchange interaction profiting the pair breaking suppresses the excitation energy $\eta\Delta_0$. The hybridization of the YSR states in ensembles of magnetic impurities at finite concentration then leads to the impurity Shiba band\cite{Shiba}, showing the finite density of states centered around $\eta\Delta_0$ with bandwidth proportional to the square root of the impurity density. On the other hand, with the increase of the magnetic impurities, the loss of the free Cooper pairs leads to a suppressed energy gap $\Delta_0$ of the Bogoliubov quasiparticle as revealed by Shiba by self-consistently solving the gap equation\cite{Shiba}.

Moreover, one can also understand the similarities in the behaviors of the phase modes in Eqs.~(\ref{pm1}) and (\ref{pm2}), as the revealed phase mode on Josephson islands in describing the granular superconductors\cite{GS1} exhibits similar behavior of the Nambu-Goldstone phase mode\cite{gi0,AK,Ba0,Am0,GIKE2}. Particularly, it is noted that for the phase mode on the state of localized Cooper pair in Eq.~(\ref{pm2}), the energy gap $\sqrt{\frac{{\rm Im}(\chi_{\rho\rho}\eta_s)ne^2}{{\rm Im}(\chi_{\rho\rho})\epsilon_0m}}=\sqrt{\frac{[{\rm Im}(\chi_{\rho\rho}){\rm Re}(\eta_s)+{\rm Re}(\chi_{\rho\rho}){\rm Im}(\eta_s)]ne^2}{{\rm Im}(\chi_{\rho\rho})\epsilon_0m}}$ at long-wavelength limit involves not only the contribution of the superfluid density ${\rm Im}(\eta_s)n$ in the localized state, but also the one ${\rm Re}(\eta_s)n$ in the free state. This is because that the electric long-range Coulomb interaction between the free and localized states is inevitable. Furthermore, we show in the following sections that the proposed picture of the ground state with free and localized Cooper pairs can well capture the obtained electromagnetic properties in conventional superconductors with magnetic impurities.

\subsection{Diamagnetic property}
\label{sec-dp}

In this part, by considering a stationary and transverse vector potential, we derive the diamagnetic response of conventional superconductors with $s$-$d$ interaction from magnetic impurities. In this situation, one has $\mu_{\rm eff}=0$ and hence $\mu_{H}=0$  as well as ${\bf p}_s=-e{\bf A}$ in Eq.~(\ref{naction}). The generated diamagnetic supercurrent from the non-equilibrium action in Eq.~(\ref{naction}) then reads
\begin{equation}\label{js}
  {\bf j}_s=-e\partial_{{\bf p}_s}\delta{S}=-\frac{\chi_v+\chi_{jj}}{2}\frac{e^2{\bf A}}{m}=-\frac{\eta_sne^2{\bf A}}{m},
\end{equation}
with the ratio of the superfluid density to the electron density from Eq.~(\ref{esn}) written as
\begin{equation}
  \eta_s=\sum_l\frac{\pi{T}\tilde\Delta_0^2}{(\tilde\Delta_0^2\!+\!\tilde\omega_l^2)^{3/2}}. \label{es}
\end{equation}

As a self-consistent check, with the vanishing renormalization (i.e., $\tilde\omega\rightarrow\omega$ and $\tilde\Delta_0\rightarrow\Delta_0$) in the absence of the magnetic impurities,  one has $\eta_s=1$ at $T=0~$K and $\eta_s=\frac{7\Delta_0^2\zeta(3)}{4(\pi{T})^2}$ near $T_c$ from Eq.~(\ref{es}), which are exactly same as the established superfluid density in the literature by various approaches\cite{G1,DDT,GIKE1,aa2}. Moreover, with the magnetic impurities, at the case above $T_c$, one has $\Delta_0=0$ and hence $\tilde\Delta_0=0$ from Eq.~(\ref{RD}), and then, the supercurrent in Eq.~(\ref{es}) vanishes, i.e., the drive current exactly cancels the pump current one in normal metals as it should be\cite{G1}, since the stationary magnetic vector potential can not drive the normal-state current even with the magnetic impurities.

With the magnetic impurities in superconductors, due to the complex renormalization by $s$-$d$ interactions, there emerges an imaginary part in the superfluid-density ratio $\eta_s$, i.e., the presence of the magnetic impurities lead to a finite imaginary part in the generated supercurrent. This imaginary part can be understood as follows based on the proposed picture of the ground state with free and localized Cooper pairs in Sec.~\ref{pp}. Specifically, between the states of the free and localized Cooper pairs, the wave-vectors exhibit a $\pi/2$-phase difference, and hence, the induced center-of-mass momenta by vector potential, which are related to the generation of the supercurrent, also have a $\pi/2$-phase difference. Therefore, in comparison with the state of the free Cooper pairs that contributes to the real part in the supercurrent, the state of the free Cooper pairs leads to an imaginary part in the supercurrent. 

It is noted that the real part in supercurrent guarantees the diamagnetic effect in the magnetic response, whereas in contrast to the conventional exponential decay at the case without magnetic impurities\cite{London}, the induced imaginary part in supercurrent due to the magnetic impurities is incapable of causing the relaxation of the supercurrent, but leads to an oscillation in the decay of the vector potential from the surface to the interior of superconductors in the diamagnetic response, similar to the Friedel oscillation in normal metals due to the local modulation of the charge density by defect\cite{MH}. Therefore, we refer to this oscillation as superconducting Friedel oscillation. Particularly, from Eq.~(\ref{js}), together with the Maxwell equation, one can obtain the equation of the vector potential, and then, solve the penetration depth $\lambda_d$ as well as the characteristic length $\lambda_o$ of the oscillation through the following equation:
\begin{equation}
\Big(\frac{1}{\lambda_d}+\frac{i}{\lambda_o}\Big)^2=\frac{4{\pi}\eta_sne^2}{m}.  
\end{equation}
At low concentration of the magnetic impurities, one has
\begin{eqnarray}
  {\lambda_d}&=&{\lambda_c}/{\sqrt{{\rm Re}\eta_s}},\label{lad}\\
{\lambda_o}&=&{2\lambda_c{\sqrt{{\rm Re}\eta_s}}}/{{\rm Im}\eta_s},
\end{eqnarray}
where $\lambda_c=\sqrt{m/{(4{\pi}ne^2)}}$ denotes the London clean-limit penetration depth at zero temperature. 

The oscillatory decay of the vector potential provides a feasible detection scheme for the involved $s$-$d$ interaction and in particular, impurity Shiba bands in superconductors with magnetic impurities, via using the muon spin relaxation ($\mu$SR) measurements\cite{MuSR}. It is also noted that the oscillatory decay in superconductors with magnetic impurities has totally different origin from the observed one in superconducting proximity structure with triplet Cooper pairs induced by magnetism\cite{MuSR}. In that case, the emerging oscillation comes from the paramagnetic Meissner effect [i.e., ${\rm Re}(\eta_s)<0$, directly leading to an imaginary $\lambda_d$ in Eq.~(\ref{lad})] by triplet Cooper pairs\cite{TC1}, and the decay is due to the suppressed gap during the diffusion in proximity structure\cite{TC2}. 

\subsection{Optical absorption}
\label{sec-oa}

We next derive the optical absorption of conventional superconductors with magnetic impurities to present a more determined detection scheme for the impurity Shiba bands. Following the Mattis-Bardeen theory\cite{MB,MBo}, we also consider a conventional $s$-wave superconductor lying in the anomalous-skin-effect region with a mean free path $L$ larger compared with the skin depth $\lambda$\cite{NSL0,GIKE3}, where the excited current at one space point depends not only on the electric field at that point but also on the ones nearby. This non-local effect in fact provides an effective dipole in the optical response, leading to the emergence of the optical absorption. The excited current in this situation reads\cite{MB,MBo}:
\begin{equation}
{\bf j}({\bf r})=\int\frac{{\bf R}[{\bf R}\cdot{\bf A}({\bf r}')]I(\Omega,{\bf R})e^{-R/L}}{R^4}d{\bf r'},  
\end{equation}
where ${\bf R}={\bf r}-{\bf r}'$; the normalized linear-response coefficient $I(\Omega,{\bf R})=\Pi(\Omega,{\bf R})/(k_F^2/3)$ with $\Pi(\Omega,{\bf R})$ denoting the linear-response coefficient and $\Omega$ representing the optical frequency. At dirty limit with a larger coherence length $\xi$ compared with the mean free path $L$ (i.e., $\xi>L>\lambda$), by the mean value theorem of integrals, one has
\begin{equation}
  {\bf j}({\bf r}){\approx}I(\Omega,{\bf R}=0){\bf A}({\bf r})\!\!\int\!\frac{e^{-R/L}}{3R^2}d{\bf r'},
\end{equation}
which leads to the optical conductivity:
\begin{equation}\label{oc}
\sigma_{s}=\sigma_{1s}+i\sigma_{2s}=\frac{4\pi{L}}{3i\Omega}\sum_{\bf q}I(\Omega,{\bf q}).  
\end{equation}  

The artificial scheme of taking the external optical frequency as imaginary bosonic Matsubara frequency within the Matsubara representation makes it hard to directly distinguish the influence (complex renormalization) of the $s$-$d$ interaction. We therefore perform the formulation within the Keldysh formalism\cite{RS}. Specifically, it is noted that the direct density-vertex contribution by pump effect in the non-equilibrium action as an unphysical non-gauge-invariant current makes no contribution to the optical absorption, and only the current-current correlation contributes to $\sigma_{1s}$ as a connected diagram in Fig.~\ref{figyw2}. Therefore, within the Keldysh space, substituting Eq.~(\ref{jj}) into Eq.~(\ref{oc}), one has
\begin{eqnarray}
  \sigma_{1s}\!\!&=&\!\!-{\rm Re}\Big[\frac{2e^2\pi{L}}{i\Omega{m}k_F^2}\sum_{\bf q}\chi_{jj}(\Omega,{\bf q})\Big]\nonumber\\
  &=&\frac{2e^2\pi{L}}{3\Omega{m^2}}\!\int\!\frac{dE}{2\pi}\!\sum_{{\bf kq}}\frac{{\rm Tr}{\rm Re}}{4}\{[{\hat G}_{\bf k^+}(E^+){\hat G}_{\bf k}(E)]_K\},~~~~
\end{eqnarray}
where ${\bf k^+}={\bf k+q}$ and ${E^+}=E+\Omega$; the subscript ``K'' denotes the Keldysh component; the Green function matrices ${\hat G}_{\bf k}(E)$ is defined as\cite{RS}
\begin{equation}
  {\hat G}_{\bf k}(E)=\left(\begin{array}{cc}
    G^R_{\bf k} & G_{\bf k}^K \\
    0 & G_{\bf k}^A
  \end{array}\right),
\end{equation}
and it is established in the literature\cite{RS,PT3,Eilen1} that the retarded~(R), advanced (A) and Keldysh (K) Green functions can be obtained by {\small $G^R_{\bf k}(E)=G_{\bf k}(E+i0^+)$, $G_{\bf k}^A(E)=G_{\bf k}(E-i0^+)$} and {\small $G^K_{\bf k}(E)=h(E)[G^R_{\bf k}(E)-G^A_{\bf k}(E)]$}, respectively, with the distribution function {\small $h(E)=\tanh(\beta{E}/2)$}.
Here, {\small $\beta=1/(k_BT)$}. 

Using the facts that {\small ${\rm Re}G^R_{\bf k}(E)={\rm Re}G^R_{\bf k}(E)$} and {\small ${\rm Im}G^A_{\bf k}(E)=-{\rm Im}G^A_{\bf k}(E)$}, the optical absorption becomes 
\begin{eqnarray}
  \sigma_{1s}\!\!&=&\frac{2e^2\pi{L}}{3\Omega{m^2}}\int\frac{dE}{2\pi}\sum_{{\bf kq}}{\rm Tr}[{\rm Im}G^R_{\bf k^+}(E^+){\rm Im}G^R_{\bf k}(E)]\nonumber\\
  &&\mbox{}\times\frac{h(E^+)\!-\!h(E)}{2},
\end{eqnarray}
and through the replacement $\sum_{\bf kq}\rightarrow\sum_{\bf kk^+}$, one obtains
\begin{eqnarray}
  \sigma_{1s}=\sigma_n\int{dE}\frac{f(E)\!-\!f(E^+)}{\Omega}\frac{m(E)\rho(E^+)\rho(E)}{\pi^2D^2},~~\label{foe}
\end{eqnarray}
where $\sigma_{n}=\frac{ne^2\tau}{m}$ represents the electrical conductivity in normal metals with $\tau$ being the momentum-relaxation time; $m(E)=1+{\tilde\Delta}_0(E){\tilde\Delta}_0(E^+)/(\tilde{E}\tilde{E}^+)$ and $m(E)L$ behaves as an effective dipole mediated by the scattering; $f(E)$ denotes the Fermi distribution function. 

As a self-consistent check, with the vanishing renormalization in the absence of the magnetic impurities, the density of states $\rho(E)$ becomes finite only when energy $|E|$ lies above the superconducting gap, and one has $\rho(E)=\pi{D}\frac{E{\rm sgn}(E)}{\sqrt{E^2-\Delta_0^2}}\theta(|E|-\Delta_0)$ from Eq.~(\ref{DOE}), with $\theta(x)$ being the step function. Then, substituting $\rho(E)$ to Eq.~(\ref{foe}), the optical absorption becomes
\begin{eqnarray}
  \frac{\sigma_{1s}}{\sigma_{n}}&=&\Big[\Big(\int^{\infty}_{\Delta_0}\!+\!\int_{-\infty}^{-\Delta_0-\Omega}\Big)-\theta(\Omega\!-\!2\Delta_0)\int^{-\Delta_0}_{\Delta_0-\Omega}\Big]\nonumber\\
  &&\mbox{}\times\frac{f(E)\!-\!f(E^+)}{\Omega}\frac{(EE^+\!+\!\Delta^2_0)dE}{\sqrt{E^2\!-\!\Delta_0^2}\sqrt{(E^+)^2\!-\!\Delta_0^2}}\nonumber\\
  &=&\Big[2\int^{\infty}_{\Delta_0}\frac{f(E)\!-\!f(E^+)}{\Omega}-\theta(\Omega\!-\!2\Delta_0)\int^{-\Delta_0}_{\Delta_0-\Omega}\nonumber\\
  &&\mbox{}\times\frac{1\!-\!2f(E^+)}{\Omega}\Big]\frac{(EE^+\!+\!\Delta^2_0)dE}{\sqrt{E^2\!-\!\Delta_0^2}\sqrt{(E^+)^2\!-\!\Delta_0^2}},~~~~~\label{MBE}
\end{eqnarray}
which exactly recovers the one from the Mattis-Bardeen theory\cite{MB,MBo}. It is noted that the first and second terms in Eq.~(\ref{MBE}) correspond to the intraband and interband transitions of the Bogoliubov quasiparticles, respectively.  As mentioned in the introduction, at $T=0~$K with only the contribution of the interband transition, the optical absorption $\sigma_{1s}(\Omega)$ vanishes when {\small $\Omega<2\Delta_0$} but becomes finite above $2\Delta_0$, leading to a crossover point at $2\Delta_0$. At finite temperature, an additional quasiparticle contribution appears below $2\Delta_0$ due to the intraband transition. 

As for the case with magnetic impurities at finite concentration, according to the proposed picture of the ground state with free and localized Cooper pairs in Sec.~\ref{pp}, the impurity Shiba bands and Bogoliubov quasiparticle continuum correspond to the excitations of the ground states of the localized and free Cooper pairs, respectively, and hence, are similar to each other. Then, based on the revealed inter- and intraband transitions of the Bogoliubov quasiparticle by the Mattis-Bardeen theory\cite{MB}, one expects the inter and intraband transitions of the impurity Shiba bands as well as all interband transitions between Bogoliubov quasiparticle and impurity Shiba bands.

Specifically, with magnetic impurities, due to the emergence of the impurity Shiba bands inside the superconducting gap, the density of states becomes finite not only above the superconducting gap but also in the Shiba-band regime of $E_b<|E|<E_t$. In this situation, considering the case at zero temperature with only the interband transition, the optical absorption becomes
\begin{widetext}
\begin{eqnarray}
  \frac{\sigma_{1s}}{\sigma_n}&=&\Big\{\theta(\Omega-2\Delta_0)\int^{-\Delta_0}_{\Delta_0-\Omega}+\theta(\Omega-\Delta_0-E_b)\int^{{\rm min}(-\Delta_0,E_t-\Omega)}_{E_b-\Omega}+\theta(\Omega-E_b-\Delta_0)\int^{-E_b}_{{\rm max}(\Delta_0-\Omega,-E_t)}\nonumber\\
  &&\mbox{}+\theta(2E_t-\Omega)\theta(\Omega-2E_b)\int^{{\rm min}(-E_b,E_t-\Omega)}_{{\rm max}(-E_t,E_b-\Omega)}\Big\}\frac{m(E)\rho(E^+)\rho(E)}{\Omega\pi^2D^2}dE.\label{ITFO}
\end{eqnarray}  
\end{widetext}
It is noted in above equation that the first term corresponds to the interband transition (channel I in Fig.~\ref{figyw1}) from the Bogoliubov quasiholes to quasielectrons, which is finite at $\Omega>2\Delta_0$, leading to a crossover at $\Omega=2\Delta_0$. The second term denotes the interband transition (channel III in Fig.~\ref{figyw1}) from the Bogoliubov quasiholes to electron-type impurity Shiba band, and the third one represents the interband transition (channel IV in Fig.~\ref{figyw1}) from the hole-type impurity Shiba band to Bogoliubov quasielectrons.  The second and third terms are symmetric and hence both are finite at $\Omega>\Delta_0+E_b$, causing a crossover at $\Omega=\Delta_0+E_b$. The forth term stands for the interband transition (channel II in Fig.~\ref{figyw1}) from the hole- to electron-type impurity Shiba bands, which is finite at $2E_t>\Omega>2E_b$ and hence leads to a resonance peak from $\Omega=2E_b$ to $\Omega=2E_t$ and centered around $2\eta\Delta_0$.

Consequently, in addition to the conventional interband transition of Bogoliubov quasiparticles as revealed by Mattis-Bardeen theory\cite{MB}, due to the emergence of the impurity Shiba bands by $s$-$d$ interaction from the magnetic impurities, at zero temperature,
there also exist the interband transitions (from hole type to electron type) between the impurity Shiba bands as well as between Bogoliubov quasiparticle and impurity Shiba bands, causing a resonance peak centered around $2\eta\Delta_0$ and a crossover at $\Delta_0+E_b$ in the optical absorption, respectively, providing clear features for the detection of the impurity Shiba bands in the optical spectroscopy. 

With increase of temperature from zero, there gradually emerge the intraband transitions inside the Bogoliubov quasiparticle continuum and inside the impurity Shiba band. Interestingly, two additional interband transtions also emerge at nonzero temperature: from electron-type impurity Shiba band to the Bogoliubov quasielectrons (channel V in Fig.~\ref{figyw1}); from the Bogoliubov quasiholes to the hold-type impurity Shiba band (channel VI in Fig.~\ref{figyw1}), leading to the contribution:
\begin{eqnarray}
  &&\!\!\!\!\!\!\!\!\!\!\frac{\sigma_{1s}}{\sigma_{n}}\Big|^{eS\rightarrow{eB}}_{hB\rightarrow{hS}}=\Big[\int^{E_t}_{{\rm max}(\Delta_0\!-\!\Omega,E_b)}\!+\!\int^{{\rm min}(-\!\Delta_0,-\!E_b\!-\!\Omega)}_{-\!E_t\!-\!\Omega}\Big]m(E)\nonumber\\
    &&\!\!\!\!\!\!\!\!\!\!\mbox{}\times\!\theta(\Omega\!+\!E_t\!-\!\Delta_0)\frac{[f(E)\!-\!f(E^+)]\rho(E^+)\rho(E)}{\Omega\pi^2D^2}dE.
\end{eqnarray}
This contribution becomes finite at $\Omega>\Delta_0-E_t$, and hence, a crossover at $\Omega=\Delta_0-E_t$ in the optical absorption gradually emerges with increase of temperature from $T=0~K$, also providing a clear feature for the detection of the impurity Shiba bands in the optical spectroscopy.

\section{Summary and Discussion}

In summary, in conventional superconductors with magnetic impurities, via analytically solving the renormalized Green function by the $s$-$d$ interaction at low impurity concentration, we have derived the macroscopic superconducting phase fluctuation, and found that there exist two superconducting phase modes. Consequently, by the Goldstone theorem\cite{Gm1,Gm2} of the collective phase mode in superconductors\cite{gi0,Gm1,Gm2,AK,Ba0,Am0,GIKE2}, we suggest that there exists a state of localized Cooper pairs around magnetic impurities, besides the one of the conventional free Cooper pairs. Then, we derived the electromagnetic properties in the linear regime to study the influence of the emerging impurity Shiba bands inside the superconducting gap.

On one hand, in the diamagnetic response, the state of the localized Cooper pairs due to the magnetic impurities results in an imaginary contribution in supercurrent, which leads to an oscillation in the decay of the vector potential in the Meissner effect, i.e., superconducting Friedel oscillation, similar to the Friedel oscillation in normal metals due to the local modulation of the charge density by defects\cite{MH}. It is noted that the superconducting Friedel oscillation is a unique character of the magnetic impurities, which breaks the time-reversal symmetry and leads to the complex renormalization, in contrast to the null renormalization\cite{ISE1,ISE2,ISE3,ISE4} (i.e., $\tilde\omega/\tilde\Delta_0=\omega/\Delta_0$) by non-magnetic impurities as the Anderson theorem revealed\cite{ISE}. Hence, this oscillation provides a feasible scheme to detect the involved $s$-$d$ interaction in superconductors with magnetic impurities, via using $\mu$SR measurement\cite{MuSR}.

On the other hand, in the optical absorption of a conventional $s$-wave superconductor lying in the anomalous-skin-effect region\cite{NSL0,GIKE3}, besides the conventional interband transition of Bogoliubov quasiparticles as revealed by Mattis-Bardeen theory\cite{MB}, there also exist the interband transitions (from hole type to electron type) between the impurity Shiba bands as well as between Bogoliubov quasiparticle and impurity Shiba bands. Additional interband transitions from hole (electron) type to hole (electron) type between Bogoliubov quasiparticle and impurity Shiba bands gradually emerge with increase of temperature. All these interband transitions give clear and separate resonance characters in the optical spectroscopy, providing a feasible scheme for the experimental detection of the impurity Shiba band. It is noted that the interband transitions between Bogoliubov quasiparticle and impurity Shiba bands are unique characters of superconductors lying in the anomalous-skin-effect region, where the excited current at one space point depends not only on the optical field at that point but also the ones nearby\cite{NSL0,GIKE3}, leading to the coupling (i.e., effective dipole) between two excitations.

\begin{acknowledgments}
The authors acknowledge financial support from
the National Natural Science Foundation of 
China under Grants No.\ 11334014 and No.\ 61411136001.  
\end{acknowledgments}

\begin{widetext}
\begin{appendix}

\section{Derivation of solution of the renormalization within real-frequency representation}
\label{sec-a1}

In this part, we present the derivation of the solution of the renormalization of $\tilde\omega/\tilde\Delta$. At low concentration of magnetic impurities, the narrow impurity Shiba band is away from the edge of the Bogoliubov quasiparticle continuum.

For the branch of the solutions of the impurity Shiba bands at $\omega>0$, the real and imaginary parts of Eq.~(\ref{ISIS1}) are written as
\begin{eqnarray}
  \delta{x}&=&r\frac{[(x\!+\!\delta{x})\sqrt{1\!-\!x^2}\!+\!m^2x/\sqrt{1\!-\!x^2}][\eta^2\!-\!(x\!+\!\delta{x})^2\!+\!m^2]\!+\!2m^2(x\!+\!\delta{x})^2x/\sqrt{1\!-\!x^2}\!-\!2m^2(x\!+\!\delta{x})\sqrt{1\!-\!x^2}}{[\eta^2\!-\!(x\!+\!\delta{x})^2\!+\!m^2]^2\!+\!4m^2(x\!+\!\delta{x})^2}, \label{dx1} \\
  m&=&mr\frac{[\sqrt{1\!-\!x^2}\!-\!(x\!+\!\delta{x})x/\sqrt{1\!-\!x^2}][\eta^2\!-\!(x\!+\!\delta{x})^2\!+\!m^2]\!+\!2(x\!+\!\delta{x})^2\sqrt{1\!-\!x^2}\!+\!2m^2(x\!+\!\delta{x})x/\sqrt{1\!-\!x^2}}{[\eta^2\!-\!(x\!+\!\delta{x})^2\!+\!m^2]^2\!+\!4m^2(x\!+\!\delta{x})^2}. \label{m1}
\end{eqnarray}
Considering the fact that the real part $\delta{x}$ of the renormalization is a small quantity compared to $x$, keeping the lowest order of $r$, the solution of $\delta{x}$ is directly given by Eq.~(\ref{R1}). Moreover, one can also neglect $\delta{x}$ in the equation of the imaginary part, and then, Eq.~(\ref{m1}) becomes
\begin{equation}
(\eta^2\!-\!x^2\!+\!m^2)^2\!+\!4m^2x^2=r\sqrt{1\!-\!x^2}(\eta^2\!-\!x^2\!+\!m^2)\!-\!rx^2/\sqrt{1\!-\!x^2}(\eta^2\!-\!x^2\!+\!m^2)\!+\!2rx^2\sqrt{1\!-\!x^2}\!+\!2rm^2x^2/\sqrt{1\!-\!x^2},  
\end{equation}
which can be re-written as
\begin{equation}
m^4+2B(x)m^2+(\eta^2-x^2)^2-rW(x)=0, 
\end{equation}
leading to the solution in Eq.~(\ref{mS}).  

Similarly, for the branch of the solutions of the continuum of the Bogoliubov quasiparticle, considering the fact that the real part $\delta{x}$ of the renormalization is a small quantity compared to $x$, the real part of Eq.~(\ref{IQIQ1}) directly becomes the solution of $\delta{x}$ in Eq.~(\ref{R2}), whereas the imaginary part reads
\begin{equation}
m=\frac{rx}{\eta^2-x^2}\sqrt{\frac{\sqrt{(x^2-1-m^2)^2+4m^2x^2}+x^2-1-m^2}{2}},
\end{equation}
and can be re-written as
\begin{equation}
\Big[1+\frac{r^2x^2}{2(\eta^2-x^2)^2}\Big]m^2-\frac{r^2x^2(x^2-1)}{2(\eta^2-x^2)^2}=\frac{r^2x^2}{2(\eta^2-x^2)^2}\sqrt{(x^2-1-m^2)^2+4m^2x^2}\approx\frac{r^2x^2(x^2-1)}{2(\eta^2-x^2)^2},
\end{equation}  
where we have kept the lowest order of $r$. Consequently, the solution of $m$ in Eq.~(\ref{mSQ}) is obtained.

\section{Derivation of solution of the renormalization within Matsubara-frequency representation}
\label{sec-a2}

In this part, we present the derivation of the solution of the renormalization within the Matsubara-frequency representation. For Matsubara frequency $\omega_l$,  by defining $x_l={\omega_l}/{\Delta_0}$, we consider a complex solution of the renormalization:
\begin{equation}
{\tilde\omega_l}/{\tilde\Delta_0}=x_l+\delta{x_l}+im_l,  
\end{equation}
in which the parameters $\delta{x_l}$ and $m_l$ are small quantities for weak renormalization at low impurity concentration.

It is noted that with $\omega\rightarrow{i\omega_l}=(2l+1)\pi{T}$, Eq.~(\ref{RD}) is unchanged, whereas Eq.~(\ref{RE}) becomes different and is written as
\begin{equation}
\frac{\omega_l}{\Delta_0}=\frac{\tilde\omega_l}{\tilde\Delta_0}\bigg[1\!-\!r\frac{\sqrt{1\!+\!(\frac{\tilde\omega_l}{\tilde\Delta_0})^2}}{\eta^2\!+\!(\frac{\tilde\omega_l}{\tilde\Delta_0})^2}\bigg],  
\end{equation}
which can be re-written as
\begin{equation}\label{A1}
\delta{x_l}+im_l=r\frac{(x_l+\delta{x_l}+im_l)\sqrt{1\!+\!(x_l+\delta{x_l}+im_l)^2}}{\eta^2\!+\!(x_l+\delta{x_l}+im_l)^2}.
\end{equation}

Considering the facts that $\delta{x}$ is a small quantity compared to $x$ and $m_l^2\ll{1+x_l^2}$ for weak renormalization at low impurity concentration, one approximately has
\begin{equation}
\delta{x_l}+im_l=r(x_l+im_l)\frac{\sqrt{1+x_l^2}+im_lx_l/\sqrt{1+x_l^2}}{[\eta^2\!+\!(x_l+im_l)^2]},
\end{equation}
which can be separated into two equations:
\begin{eqnarray}
  (\eta^2+x_l^2-m_l^2)\delta{x}_l-2x_lm_l^2=rx_l\sqrt{1+x_l^2}-{rm^2_lx_l}{\sqrt{1+x_l^2}},\label{BB1} \\
  2x_l\delta{x_l}=r\sqrt{1+x_l^2}+{rx^2_l}/{\sqrt{1+x_l^2}}+m_l^2-\eta^2-x_l^2.\label{BB2}
\end{eqnarray}

By keeping the lowest two orders of $r$ and solving Eqs.~(\ref{BB1}) and (\ref{BB2}), at $\omega_l>0$, one finds the solutions:
\begin{eqnarray}
  2x_l\delta{x}=\frac{r/2}{\sqrt{1+x_l^2}}+\frac{2rx_l^2}{\sqrt{1+x_l^2}}-2x_l^2+2ix_l\eta\Big(1-\frac{r/4}{\sqrt{1+x_l^2}}\Big)\\
  m=-\eta\Big(1-\frac{r}{4}\frac{\sqrt{1+x_l^2}}{\eta^2+x_l^2}\Big)-ix_l\Big[1+\frac{r}{4}\frac{1-\eta^2}{\sqrt{1+x_l^2}(\eta^2+x_l^2)}\Big]
\end{eqnarray}  
and hence,
\begin{equation}\label{FAA}
\Big(\frac{\tilde\omega_l}{\tilde\Delta_0}\Big)^2=x_l^2+2x_l(\delta{x}_l+im_l)=\frac{r}{2}\Big[\frac{1+4x_l^2}{\sqrt{1+x_l^2}}+\frac{x^2_l(1-\eta^2)}{\sqrt{1+x_l^2}(\eta^2+x_l^2)}\Big]+\frac{i{\eta}r}{2}\frac{x_l}{\eta^2+x_l^2}\frac{1-\eta^2}{\sqrt{1+x_l^2}}.
\end{equation}
Consequently, differing from the solution in the real-frequency representation as obtained in Sec.~\ref{sec-a1}, the solution of the renormalization by the $s$-$d$ interaction in the Matsubara-frequency representation is always complex. As a self-consistent check, in the case without magnetic impurities ($r=\rightarrow0$), the renormalization in Eq.~(\ref{FAA}) vanishes as it should be. Moreover, due to the factor $x_l/(\eta^2+x_l^2)$, the imaginay part of the renormalization in Eq.~(\ref{FAA}), which is related to the contribution from the state of the localized Cooper pairs, achieves the maximum at $x_l=\eta$. Consequently, as the minimum of $x_l$ is $\pi{T}/\Delta_0$, the increase of temperature at ${\pi}T>\eta\Delta_0$ leads to the suppressions on the imaginay part of the renormalization and hence the imaginary part of the superfluid-density ratio $\eta_s$ [Eq.~(\ref{esn})], suggesting the breaking of the localized Cooper pairs by the excitation of the YSR state. By further increasing temperature until $x_l\gg{\eta}$ at all $l$, the imaginay part of the renormalization nearly vanishes due to the vanishing localized Cooper pairs. \\

\end{appendix}

\end{widetext}


\begin{thebibliography}{0}


\bibitem{VB1} A. A. Abrikosov, Zh. Eksp. Teor. Fiz. {\bf 32}, 1442 (1957).
\bibitem{VB2} C. Caroli, P. G. de Gennes, and J. Matricon, Phys. Lett. {\bf 9}, 307 (1964).

\bibitem{Yu} L. Yu, Acta. Phys. Sin. {\bf 21}, 75 (1965).
\bibitem{Shiba} H. Shiba, Prog. Theor. Phys. {\bf 40}, 435 (1968).
\bibitem{Ru} A. I. Rusinov, JETP Lett. {\bf 9}, 85 (1969).

\bibitem{AB1} P. G. de Gennes and D. Saint-James,  Phys. Lett. {\bf 4}, 151 (1963).
\bibitem{AB2} I. O. Kulik, Sov. Phys. JETP {\bf 30}, 944 (1970).
\bibitem{AB3} G. E. Blonder, M. Tinkham, and T. M. Klapwijk, Phys. Rev. B {\bf 25}, 4515 (1982).
\bibitem{AB4} G. Kells, D. Meidan, and P. Brouwer, Phys. Rev. B {\bf 86}, 100503 (2012).
\bibitem{AB5} C.-X. Liu, J. D. Sau, T. D. Stanescu, and S. D. Sarma, Phys. Rev. B {\bf 96}, 075161 (2017).
\bibitem{AB6} J. A. Sauls,  Phil. Trans. R. Soc. A {\bf 376}, 20180140 (2018).

\bibitem{MB1}  A. Y. Kitaev, Phys. Uspekhi {\bf 44}, 131 (2001).
\bibitem{MB2}  C. Nayak, S. H. Simon, A. Stern, M. Freedman, and S. Das Sarma, Rev. Mod. Phys. {\bf 80}, 1083 (2008).
\bibitem{MB3} L. Fu and C. L. Kane, Phys. Rev. Lett. {\bf 100}, 096407 (2008).
\bibitem{MB4} R. M. Lutchyn, J. D. Sau, and S. Das Sarma, Phys. Rev. Lett. {\bf 105}, 077001 (2010).
\bibitem{MB5}  Y. Oreg, G. Refael, and F. von Oppen, Phys. Rev. Lett. {\bf 105}, 177002 (2010).
\bibitem{MB6}  C. W. J. Beenakker, Annu. Rev. Condens. Matter Phys. {\bf 4}, 113 (2013).
\bibitem{MB7} M. T. Deng, S. Vaitiekenas, E. B. Hansen, J. Danon, M. Leijnse, K. Flensberg, J. Nygard, P. Krogstrup, and C. M. Marcus, Science {\bf 354}, 1557 (2016).


  
\bibitem{SA} M. Ruby, F. Pientka, Y. Peng, F. von Oppen, B. W. Heinrich, and K. J. Franke, Phys. Rev. Lett. {\bf 115}, 087001 (2015). 

\bibitem{SM1} R. Zitko, O. Bodensiek, and T. Pruschke, Phys. Rev. B {\bf 83}, 054512 (2011).
\bibitem{SM2} N. Hatter,  B. W. Heinrich, M. Ruby, J. I. Pascual, and K. J. Franke, Nat.
Commun. {\bf 6}, 8988 (2015). 
\bibitem{SM3} G. C. M{\'e}nard, S. Guissart, C. Brun, S. Pons, V. S. Stolyarov, F. Debontridder, M. V. Leclerc, E. Janod, L. Cario, D. Roditchev, P. Simon, and T. Cren, Nat. Phys. {\bf 11}, 1013 (2015).
\bibitem{SM4}  E. W. Hudson, K. M. Lang, V. Madhavan, S. H. Pan, H. Eisaki, S. Uchida, and J. C. Davis. Nature {\bf 411}, 920 (2001).
  

\bibitem{SAP1} T.-P. Choy, J. M. Edge, A. R. Akhmerov, and C. W. J. Beenakker, Phys. Rev. B {\bf 84}, 195442 (2011).
\bibitem{SAP2} J. Klinovaja, P. Stano, A. Yazdani, and D. Loss, Phys. Rev. Lett. {\bf 111}, 186805 (2013).
\bibitem{SAP3}  S. Nadj-Perge, I. K. Drozdov, B. A. Bernevig, and A. Yazdani,  Phys. Rev. B {\bf 88}, 020407 (2013).  
\bibitem{SAP4} S. Nadj-Perge, I. K. Drozdov, J. Li, H. Chen, S. Jeon, J. Seo,
A. H. MacDonald, B. A. Bernevig, and A. Yazdani, Science {\bf 346}, 602 (2014).
\bibitem{SAP5} K. P{\"o}yh{\"o}nen, I. Sahlberg, A. Weststr{\"o}m, and T. Ojanen, Nat. Commun. {\bf 9}, 2103 (2018).
\bibitem{SAP6} H. Kim, A. Palacio-Morales, T. Posske, L. R{\'o}zsa, K. n. Palot{\'a}s,
  L. S. Szunyogh, M. Thorwart, and R. Wiesendanger, Sci. Adv. {\bf 4}, eaar5251 (2018).
\bibitem{SAP7} D. Sticlet and C. Morari, Phys. Rev. B {\bf 100}, 075420 (2019).
\bibitem{SAP8} L. Schneider, S. Brinker, M. Steinbrecher, J. Hermenau, T. Posske, M. D. S. Dias, S. Lounis, R. Wiesendanger, and J. Wiebe, Nat. Commun. {\bf 11}, 4707 (2020).
\bibitem{SAP9} L. Schneider, P. Beck, T. Posske, D. Crawford, E. Mascot, S. Rachel, R. Wiesendanger, and J. Wiebe, Nat. Phys. {\bf 17}, 943 (2021).



\bibitem{STM0} A. Yazdani, B. A. Jones, C. P. Lutz, M. F. Crommie, and
D. M. Eigler, Probing the local effects of magnetic impurities
on superconductivity, Science 275, 1767 (1997). 


\bibitem{STMA1} S.-H. Ji, T. Zhang, Y.-S. Fu, X. Chen, X.-C. Ma, J. Li,
  W.-H. Duan, J.-F. Jia, and Q.-K. Xue, Phys. Rev. Lett. {\bf 100}, 226801 (2008). 
\bibitem{STMA2} D.-J. Choi, C. Rubio-Verd{/'u}, J. de Bruijckere, M. M. Ugeda,
N. Lorente, and J. I. Pascual,  Nat. Commun. {\bf 8},
15175 (2017). 
\bibitem{STMA3} M. Ruby, Y. Peng, F. von Oppen, B. W. Heinrich, and K. J.
Franke, Phys. Rev. Lett. {\bf 117}, 186801 (2016). 
\bibitem{STMA4} A. Odobesko, D. Di Sante, A. Kowalski, S. Wilfert, F. Friedrich,
  R. Thomale, G. Sangiovanni, and M. Bode, Phys. Rev. B {\bf 102}, 174504 (2020). 
\bibitem{STMA5} S. Y. Song, Y. S. Park, Y. Jeong, M.-S. Kim, K.-S. Kim, and
  J. Seo, Phys. Rev. B {\bf 103}, 214509 (2021). 
\bibitem{STMA6} F. K{\"u}ster, S. Brinker, S. Lounis, S. S. P. Parkin, and P. Sessi, Nat. Commun. {\bf 12}, 6722 (2021). 
\bibitem{STMA7} F. Friedrich, R. Boshuis, M. Bode, and A. Odobesko, Phys. Rev. B {\bf 103}, 235437 (2021). 
\bibitem{STMA8} P. Beck, L. Schneider, L. R\'ozsa, K. Palot\'as, A. L\'aszl\'offy, L. Szunyogh, J. Wiebe, and R. Wiesendanger, Nat. Commun. {\bf 12}, 2040 (2021). 


  
\bibitem{STMM1} K. J. Franke, G. Schulze, and J. I. Pascual, Science {\bf 332}, 940 (2011). 
\bibitem{STMM2} L. Malavolti, M. Briganti, M. H\"anze, G. Serrano, I. Cimatti,
G. McMurtrie, E. Otero, P. Ohresser, F. Totti, M. Mannini, R.
Sessoli, and S. Loth, Nano Lett. {\bf 18}, 7955 (2018).
\bibitem{STMM3} J. Brand, S. Gozdzik, N. N\'eel, J. L. Lado, J. Fern\'andez-Rossier,
and J. Kr\"oger, Phys. Rev. B {\bf 97}, 195429 (2018).
\bibitem{STMM4} S. Kezilebieke, M. Dvorak, T. Ojanen, and P. Liljeroth, Nano
Lett. {\bf 18}, 2311 (2018).
\bibitem{STMM5} S. Kezilebieke, R. Zitko, M. Dvorak, T. Ojanen, and P. Liljeroth, Nano Lett. {\bf 19}, 4614 (2019). 


\bibitem{STMI1} A. Palacio-Morales, E. Mascot, S. Cocklin, H. Kim, S. Rachel,
D. K. Morr, and R. Wiesendanger, Sci. Adv. {\bf 5}, eaav6600 (2019).
\bibitem{STMI2} S. Kezilebieke, M. N. Huda, V. Vano, M. Aapro, S. C. Ganguli, O. J. Silveira, S. Glodzik, A. S. Foster, T. Ojanen, and P. Liljeroth, Nature {\bf 588}, 424 (2020).

  
\bibitem{STMP} J. O. Island, R. Gaudenzi, J. de Bruijckere, E. Burzur\'i, C. Franco, M. M. Torrent, C. Rovira, J. Veciana, T. M. Klapwijk, R. Aguado, and H. S. J. van der Zant, Phys. Rev. Lett. {\bf 118}, 117001 (2017).

  
\bibitem{STMHT1} S. Y. Song, J. H. J. Martiny, A. Kreisel, B. M. Andersen, and J.
Seo, Phys. Rev. Lett. {\bf 124}, 117001 (2020).
\bibitem{STMHT2} D. Wang, J. Wiebe, R. Zhong, G. Gu, and R. Wiesendanger, Phys. Rev. Lett. {\bf 126}, 076802 (2021). 
\bibitem{STMHT3} D. Chatzopoulos, D. Cho, K. M. Bastiaans, G. O. Steffensen,
D. Bouwmeester, A. Akbari, G. Gu, J. Paaske, B. M. Andersen,
and M. P. Allan, Nat. Commun. {\bf 12}, 298 (2021). 

\bibitem{Meissner} W. Meissner and R. Ochsenfeld, Naturwissenschaften {\bf 21},
  787 (1933). 
\bibitem{London}  F. London and H. London, Proc. R. Soc. A {\bf 149}, 71 (1935).

  
\bibitem{NSL0} M. Dressel, Adv. Condens. Matter Phys. {\bf 2013}, 104379. 
\bibitem{GIKE3} F. Yang and M. W. Wu,  Phys. Rev. B {\bf 102}, 144508 (2020).

  
\bibitem{Infrad2} E. Uykur, K. Tanaka, T. Masui, S. Miyasaka, and S. Tajima,
Phys. Rev. Lett. {\bf 112}, 127003 (2014). 
\bibitem{Infrad3} K. Lee, K. Kamiya, M. Nakajima, S. Miyasaka, and S. Tajima,
  J. Phys. Soc. Jpn. {\bf 86}, 023701 (2017). 
\bibitem{MS1} H. Kitano, T. Ohashi, A. Maeda, and I. Tsukada, Phys. Rev. B
{\bf 73}, 092504 (2006). 
\bibitem{MS3} M. S. Grbi{\'c}, M. Po{\v z}ek, D. Paar, V. Hinkov, M. Raichle, D. 
Haug, B. Keimer, N. Bari{\v s}i{\'c}, and A. Dul{\v c}i{\'c}, Phys. Rev. B {\bf 83},
144508 (2011). 
\bibitem{THZ3} L. S. Bilbro, R. V. Aguilar, G. Logvenov, O. Pelleg, I. Bo{\v z}ovi{\'c}, and N. P. Armitage, Nat. Phys. {\bf 7}, 298 (2011). 
\bibitem{THZ4} D. Nakamura, Y. Imai, A. Maeda, and I. Tsukada, J. Phys. Soc.
  Jpn. {\bf 81}, 044709 (2012). 
\bibitem{NSB2} F. Mahmood, X. He, I. Bo{\v z}ovi{\'c}, and N. P. Armitage, Phys. Rev. Lett. {\bf 122}, 027003  (2019).


\bibitem{MB} D. C. Mattis and J. Bardeen, Phys. Rev. {\bf 111}, 412 (1958).
\bibitem{MBo} S. B. Nam, Phys. Rev. {\bf 156}, 470 (1967); I. S. B. Nam, Phys. Rev. B {\bf 2}, 3812 (1970).

\bibitem{L1}  N. M. Rugheimer, A. Lehoczky, and C. V. Briscoe,
Phys. Rev. {\bf 154}, 414 (1967).  
\bibitem{L2} L. H. Palmer and M. Tinkham, Phys. Rev. {\bf 165}, 588 (1968).
\bibitem{L3} D. R. Karecki, G. L. Carr, S. Perkowitz, D. U. Gubser, and S. A. Wolf,
Phys. Rev. B {\bf 27}, 5460 (1983). 
\bibitem{L4}  D. E. Oates, A. C. Anderson, C. C. Chin, J. S. Derov,
  G. Dresselhaus, and M. S. Dresselhaus, Phys. Rev. B {\bf 43}, 7655 (1991). 
\bibitem{L5} M. C. Nuss, K. W. Goossen, J. P. Gordon, P. M. Mankiewich, M. L. O'Malley, and M. Bhushan, J. Appl. Phys.
  {\bf 70}, 2238 (1991). 
\bibitem{L6} J. F. Federici, B. I. Greene, P. N. Saeta, D. R. Dykaar, F. Sharifi, and R. C. Dynes,
Phys. Rev. B {\bf 46}, 11153 (1992). 
\bibitem{L7} G. L. Carr, R. P. S. M. Lobo, J. LaVeigne, D. H. Reitze, and D. B. Tanner,
Phys. Rev. Lett. {\bf 85}, 3001 (2000). 
\bibitem{L8} K. Steinberg, M. Scheffler, and M. Dressel, Phys. Rev. B
{\bf 77}, 214517 (2008). 


\bibitem{G1} A. A. Abrikosov, L. P. Gor'kov, and I. E. Dzyaloshinski, {\em Methods of Quantum Field Theory
in Statistical Physics} (Prentice Hall, Englewood Cliffs, 1963). 
\bibitem{MH} G. D. Mahan, {\em Many Particle Physics} (Plenum, New York, 1990).

  
\bibitem{AHM} P. W. Anderson, Phys. Rev. {\bf 130}, 439 (1963).



\bibitem{gi0} Y. Nambu, Phys. Rev. {\bf 117}, 648 (1960).


  
\bibitem{AK} V. Ambegaokar and L. P. Kadanoff,  Nuovo Cimento {\bf 22}, 914
  (1961). 
\bibitem{Ba0} J. R. Schrieffer, {\em Theory of Superconductivity}
  (W. A. Benjamin, New York, 1964).
\bibitem{Am0} P. B. Littlewood and C. M. Varma, Phys. Rev. Lett. {\bf 47}, 811
(1981); Phys. Rev. B {\bf 26}, 4883 (1982). 
\bibitem{GIKE2} F. Yang and M. W. Wu,  Phys. Rev. B {\bf 100}, 104513 (2019).  
  
\bibitem{Gm1} J. Goldstone, Nuovo Cimento {\bf 19}, 154 (1961).
\bibitem{Gm2} J. Goldstone, A. Salam, and S. Weinberg, Phys. Rev. {\bf 127}, 965 (1962).





  
\bibitem{GS1}  Matthew P. A. Fisher and G. Grinstein, Phys. Rev. Lett. {\bf 60}, (1988). 
\bibitem{GS2} U. S. Pracht, N. Bachar, L. Benfatto, G. Deutscher, E. Farber, M. Dressel, and M. Scheffler, Phys. Rev. B {\bf 93}, 100503 (2016).

  
\bibitem{PT1} F. Yang and M. W. Wu, Phys. Rev. B {\bf 106}, 144509 (2022). 



\bibitem{OD1} A. F. Volkov and S. M. Kogan, Zh. Eksp. Teor. Fiz {\bf 65}, 2038
  (1974) [Sov. Phys. JETP {\bf 38}, 1018 (1974)].
\bibitem{OD2} E. A. Yuzbashyan and M. Dzero, Phys. Rev. Lett {\bf
  96}, 230404 (2006).    
\bibitem{OD3} V. Gurarie, Phys. Rev. Lett. {\bf 103}, 075301 (2009).
\bibitem{Am12} N. Tsuji and H. Aoki, Phys. Rev. B {\bf 92}, 064508 (2015).  
\bibitem{Am6} D. Pekker and C. Varma, Annu. Rev. Condens. Matter Phys. {\bf 6},
  269 (2015). 
\bibitem{symmetry} S. Tsuchiya, D. Yamamoto, R. Yoshii, and M. Nitta,  Phys. Rev. B {\bf
    98}, 094503 (2018). 

  

\bibitem{GIKE4} F. Yang and M. W. Wu,  Phys. Rev. B {\bf 102}, 014511 (2020).
\bibitem{PT2} F. Yang and M. W. Wu, Ann. Phys. {\bf 453}, 169312 (2023). 
\bibitem{PT3}  F. Yang and M. W. Wu, arXiv:2301.04832.

\bibitem{RS} J. Rammer and H. Smith, Rev. Mod. Phys. {\bf 58}, 323 (1986).
    
\bibitem{DDT} F. Yang and M. W. Wu, Phys. Rev. B {\bf 104}, 214510 (2021). 
  
\bibitem{GIKE1} F. Yang and M. W. Wu,  Phys. Rev. B {\bf 98}, 094507 (2018).
\bibitem{aa2} Z. Y. Sun, M. M. Fogler, D. N. Basov, and A. J. Millis, Phys. Rev. Research {\bf 2}, 023413
  (2020).
  
\bibitem{MuSR}  A. Di Bernardo, Z. Salman, X. L. Wang, M. Amado, M. Egilmez, M. G. Flokstra, A. Suter, S. L. Lee, J. H. Zhao, T. Prokscha, E. Morenzoni, M. G. Blamire, J. Linder, and J. W. A. Robinson, Phys. Rev. X {\bf 5}, 041021 (2015).
\bibitem{TC1} C. Espeday, T. Yokoyama, and J. Linder, Phys. Rev. Lett. {\bf 116}, 127002 (2016)
\bibitem{TC2}  F. S. Bergeret and I. V. Tokatly, Phys. Rev. Lett. {\bf 110}, 117003 (2013).

 \bibitem{Eilen1} H. G. Hugdal, J. Linder, and S. H. Jacobsen, Phys.
  Rev. B {\bf 95}, 235403 (2017).
  
\bibitem{ISE1} A. A. Abrikosov and L. P. Gor'kov, Zh. Eksp. Teor. Fiz. {\bf 35}, 1558 (1958) [Sov. Phys. JETP {\bf 8}, 1090 (1959)]; Zh. Eksp. Teor. Fiz. {\bf 36}, 319 (1959) [Sov. Phys. JETP {\bf 9}, 220 (1959)].

\bibitem{ISE2} H. Suhl and B. T. Matthias, Phys. Rev. {\bf 114}, 977 (1959).
\bibitem{ISE3} S. Skalski, O. B. Matibet, and P. R. Weiss, Phys. Rev. {\bf 136}, A1500 (1964).
\bibitem{ISE4} L. Andersen, A. Ramires, Z. W. Wang, T. Lorenz, and Y. Ando, Sci. Adv. {\bf 6}, eaay6502
(2020).


\bibitem{ISE} P. W. Anderson, J. Phys. Chem. Solids {\bf 11}, 26 (1959).



\end{thebibliography}
\end{document}